\documentclass[aps,pra,onecolumn,superscriptaddress,floatfix]{revtex4-2}

\usepackage{amsmath,amssymb,amsfonts,mathtools}
\usepackage{amsthm}
\usepackage{graphicx}
\usepackage[section]{placeins}
\usepackage{bm}
\usepackage[dvipsnames,svgnames]{xcolor}
\usepackage[colorlinks=true]{hyperref}
\usepackage{qcircuit}
\usepackage{float}
\hypersetup{
	linkcolor   = NavyBlue,
	citecolor   = BrickRed,
	urlcolor    = Teal,
	pdftitle    = {Closed Quantum Boltzmann Bridges: Coherent Revivals, Hidden Microstates, and the Emergence of Classical Two-Time Entropy Conditioning},
	pdfauthor = {Sina Kazemian, Ghazal Farhani, and Younes Javanmard},
pdfkeywords = {Quantum Boltzmann Bridges, Boltzmann entropy, two-time conditioning, quantum coarse-graining, hidden microstates, coherent revivals, Random Forest}
}

\begin{document}
	
\title{Closed Quantum Boltzmann Bridges: Coherent Revivals, Hidden Microstates, and the Emergence of Classical Two-Time Entropy Conditioning}

\author{Sina Kazemian}
\email{skazemi5@uwo.ca}
\affiliation{University of Western Ontario, Department of Physics and Astronomy, London, Ontario, Canada}
    
\author{Ghazal Farhani}
\email{ghazal.farhani@nrc-cnrc.gc.ca}
\affiliation{National Research Council Canada, London, Ontario, Canada}

\author{Younes Javanmard}
\email{javanmard.younes@gmail.com}
\affiliation{Institut f\"ur Physik und Astronomie, Technische Universit\"at Berlin, Hardenbergstra\ss e 36, EW 7-1, 10623 Berlin, Germany}
	
%\date{\today}
\begin{abstract}
The classical Boltzmann Bridge describes entropy histories conditioned on both an initial low-entropy macrostate and a later macrostate. Unlike the usual past-only formulation of the thermodynamic arrow, this two-time conditioning can produce entropy profiles that rise above the final entropy and then decrease toward the imposed endpoint. In this work, we formulate closed quantum analogues of the Boltzmann Bridge using macro-subspace projectors, unitary time evolution, and Boltzmann entropy defined by the dimension of coarse-grained macroscopic sectors. We first study a minimal coherent chamber-qubit model, in which each particle has only a two-state chamber degree of freedom. Although this model is the most direct quantization of the classical two-box system, its bridge entropy profile is dominated by coherent oscillations and revivals rather than classical relaxation. We then introduce a hidden-microstate bridge, in which each chamber sector contains unresolved internal degrees of freedom while the full dynamics remain unitary. Numerical experiments show that increasing the internal Hilbert-space dimension suppresses sample-dependent revival behavior and produces bridge entropy profiles whose sign structure and coarse-grained shape increasingly agree with the classical Boltzmann Bridge. We further use a Random Forest classifier to explore the parameter regime separating revival-dominated quantum behavior from classical-like coarse-grained bridge behavior. These results suggest that classical two-time-conditioned entropy behavior is not recovered by quantizing the chamber variable alone, but can emerge statistically from closed quantum dynamics when sufficiently rich hidden internal structure is present.
\end{abstract}

\maketitle

\section{Introduction}

Entropy occupies a central place in statistical mechanics because it connects microscopic dynamics with macroscopic irreversibility \cite{PathriaBeale2011,BrownMyrvoldUffink2009}. At the microscopic level, the fundamental laws are often reversible, while at the macroscopic level physical systems appear to evolve toward equilibrium. In the Boltzmannian view, this apparent irreversibility is explained by the fact that high-entropy macrostates correspond to overwhelmingly larger sets of microscopic configurations than low-entropy macrostates. A system prepared in a low-entropy macrostate is therefore expected, with high probability, to evolve toward macrostates of larger entropy. The same statistical-mechanical perspective also underlies nonequilibrium and transport theory, from linear-response formulations and electron--phonon transport to modern studies of phonon-mediated thermal conductivity in bulk, low-dimensional, and thin-film systems \cite{Kubo1957,Ziman1960,Broido2007,kazemian2026electron,cahill2000interface,Kazemian2019ThinFilm}.

A common explanation for the thermodynamic arrow of time is the Past Hypothesis: the assumption that the system, or the universe as a whole, began in a very low-entropy macrostate \cite{Albert2000,CarrollChen2004,Wallace2011,Chen2020}. If one conditions only on this low-entropy past, then the expected entropy typically increases toward the future. However, this argument becomes more subtle when one also conditions on a later or present macrostate. In probability theory, expectations should be conditioned on all available information \cite{Cox1946,Savage1954,DeFinetti1992}. Therefore, if both an initial low-entropy macrocondition and a later macrocondition are specified, the relevant entropy history is not simply a forward relaxation curve from the past. It is instead a two-time-conditioned entropy history.

This observation was recently developed in the classical setting under the name of the Boltzmann Bridge \cite{ScharnhorstWolpertRovelli2024}. The construction is analogous to a Brownian bridge: instead of studying unconstrained stochastic trajectories starting from one endpoint, one studies trajectories conditioned on both an initial and a final condition. In the classical Boltzmann Bridge, the system is modeled as a Markov process over macrostates, and the expected Boltzmann entropy at an intermediate time is computed by conditioning on both endpoint macrostates. This two-time conditioning can significantly alter the entropy profile. In particular, the expected entropy may rise above the entropy of the final macrostate at intermediate times and then decrease toward the imposed endpoint. Thus, the Past Hypothesis alone does not determine the expected entropy direction once present or final macroscopic information is also included.

The existing Boltzmann Bridge framework is classical and stochastic. It relies on Markov transition probabilities, Bayesian conditioning, and classical counting of microstates. A natural open question is how this construction should be extended to quantum systems. This question is nontrivial because closed quantum systems evolve unitarily rather than stochastically. Moreover, quantum macrostates are represented by subspaces and projectors rather than by classical subsets of phase space. A quantum version of the Boltzmann Bridge must therefore clarify how macroscopic conditioning, coarse-graining, and unitary time evolution interact. This also requires distinguishing Boltzmann entropy, defined by macrostate multiplicity, from von Neumann entropy, which is a property of the quantum state itself \cite{VonNeumann1927,Wehrl1978,GoldsteinLebowitzTumulkaZanghi2006,PopescuShortWinter2006}.

In this work, we formulate closed quantum analogues of the Boltzmann Bridge. We keep the same conceptual ingredients as in the classical problem: a low-entropy initial macrocondition, a final macrocondition, a macroscopic observable, and a Boltzmann entropy associated with the size of the corresponding macro-subspace. The classical macrostates are replaced by macro-subspace projectors, while the underlying microscopic dynamics remain unitary. The resulting bridge probabilities are probabilities for coarse-grained macro-histories: they involve conditioning on an intermediate macroscopic sector and then imposing the final macrocondition. In this sense, the construction is a quantum analogue of the classical two-time-conditioned entropy problem, but with the important difference that macroscopic conditioning is implemented through projectors on Hilbert-space sectors. This interpretation is closely related to the broader logic of quantum histories, postselection, and coarse-grained quantum probabilities \cite{AharonovBergmannLebowitz1964,Griffiths1984,Omnes1992,GellMannHartle1993}.

We study two quantum realizations. The first is a minimal coherent chamber-qubit model, in which each particle is represented only by a two-state chamber degree of freedom. This is the most direct quantization of the classical left-right model. However, because the dynamics are fully coherent, the resulting entropy profile exhibits oscillations and revivals rather than classical relaxation. This model therefore serves as a useful baseline, but not as a classical limit.

The second realization introduces hidden internal microstates inside each chamber sector. In this hidden-microstate bridge, the coarse chamber label remains the observed macroscopic variable, but each chamber contains unresolved internal degrees of freedom. The full dynamics remain closed and unitary, while the internal states can scramble phases when viewed through the coarse macroscopic observable. This allows us to study whether classical-looking bridge behavior can arise from closed quantum dynamics through coarse-graining, rather than through externally imposed dissipation or stochasticity.

We also use a Random Forest classifier as an exploratory tool for mapping the parameter space of quantum Boltzmann Bridges. Machine-learning methods are now widely used across the physical sciences, including quantum physics, atmospheric and geophysical modeling, astronomy, remote sensing, transport forecasting, and uncertainty-aware differential-equation-based prediction \cite{Carleo2019MachineLearningPhysics,Arcomano2020MLForecast,MarquezNeila2018ExoplanetML,Feng2019RandomForestAirPollution,Farhani2021LidarClassification,Farhani2023BayesianLidar,KazemianFarhaniYazdi2025BlackScholesPINN}. Here, the classifier is trained to distinguish classical-like bridge behavior from revival-dominated or mixed quantum behavior using model parameters and simple spectral features of the internal unitary dynamics. This analysis helps identify which combinations of endpoint constraint, internal Hilbert-space dimension, and internal spectral structure favor classical-like coarse-grained bridge behavior.

Our results show that the minimal coherent quantum model does not reproduce the classical Boltzmann Bridge. Instead, its bridge entropy is dominated by unitary revivals. By contrast, the hidden-microstate model moves progressively closer to the classical bridge pattern as the internal Hilbert-space dimension increases and the internal dynamics become more generic. This suggests that the classical Boltzmann Bridge is not recovered by quantizing the chamber variable alone. Rather, classical-like two-time-conditioned entropy behavior can emerge statistically from closed quantum dynamics when sufficiently rich unresolved internal structure is present.

\section{Classical Boltzmann Bridges}

The classical Boltzmann Bridge was introduced as a refinement of the usual Past-Hypothesis account of the thermodynamic arrow of time \cite{Albert2000,Wallace2011,Chen2020,ScharnhorstWolpertRovelli2024}. The standard argument considers the expected entropy at a later time conditioned only on a low-entropy past macrostate. In that case, the relevant object is a one-time-conditioned expectation. However, if one also specifies a later or present macrostate, probability theory requires conditioning on that information as well \cite{Cox1946,Savage1954,DeFinetti1992}. The resulting entropy curve is no longer simply a relaxation curve from the past, but a two-time-conditioned bridge.

The key conceptual point is that the usual Second-Law argument is a one-time-conditioned argument. It says that if a system begins in a low-entropy macrostate, then its entropy is expected to increase toward the future. The Boltzmann Bridge construction shows why this conclusion is incomplete without specifying the conditioning information. If a later or present macrostate is also imposed, then the expected entropy history must be conditioned on both endpoints. The typical path can then behave like a bridge: it starts from a low-entropy macrostate, rises toward higher entropy at intermediate times, and bends back toward the imposed final macrostate. This is the reason for the term \emph{Boltzmann Bridge}, in direct analogy with a Brownian bridge, where a stochastic path is conditioned on both its starting point and its ending point \cite{ScharnhorstWolpertRovelli2024}.

Let $X_t$ denote a classical macrostate evolving as a discrete-time Markov process. Suppose that the system is conditioned on an initial macrostate $X_0=x_0$ and a final macrostate $X_{t_f}=x_f$. For an intermediate time $0<t<t_f$, the Markov property and Bayes' theorem give the bridge distribution
\begin{equation}
\begin{aligned}
P(X_t=x\mid X_0=x_0,X_{t_f}=x_f)
&=
\frac{
P(X_t=x\mid X_0=x_0),
P(X_{t_f}=x_f\mid X_t=x)
}{
P(X_{t_f}=x_f\mid X_0=x_0)
}.
\end{aligned}
\label{eq}
\end{equation}
This formula is the basic structure of the classical Boltzmann Bridge. The first factor describes the forward evolution from the initial macrostate to the intermediate macrostate. The second factor measures how compatible that intermediate macrostate is with the imposed final condition. The denominator normalizes the distribution over all possible intermediate macrostates.

The Boltzmann entropy associated with a macrostate $X$ is defined by the number of microscopic states compatible with it,
\begin{equation}
S_B(X)=\log \Omega(X),
\label{eq}
\end{equation}
where $\Omega(X)$ is the macrostate multiplicity. Throughout this work, $\log$ denotes the natural logarithm. This is the Boltzmann entropy of a macrostate, not the Shannon entropy of a probability distribution over macrostates \cite{PathriaBeale2011,BrownMyrvoldUffink2009}. The bridge-conditioned expected entropy is therefore
\begin{equation}
\begin{aligned}
E[S_B(t)\mid X_0=x_0,X_{t_f}=x_f]
&=
\sum_x
S_B(x),
P(X_t=x\mid X_0=x_0,X_{t_f}=x_f).
\end{aligned}
\label{eq3}
\end{equation}
This should be compared with the past-only expectation
\begin{equation}
\begin{aligned}
E[S_B(t)\mid X_0=x_0]
&=
\sum_x
S_B(x),
P(X_t=x\mid X_0=x_0).
\end{aligned}
\label{eq4}
\end{equation}
The difference between Eqs.~\eqref{eq3} and \eqref{eq4} is central. The past-only curve asks what is expected after a low-entropy initial condition. The bridge curve asks what is expected at intermediate times, given both the low-entropy initial condition and the later macrostate.

The two-box gas model gives a concrete realization of this idea. Consider \(N\) distinguishable particles in two chambers. Each particle is represented by a classical bit \(b_i(t)\in\{0,1\}\), where \(b_i(t)=1\) means that particle \(i\) is in the left chamber and \(b_i(t)=0\) means that it is in the right chamber. The macrostate is the number of particles in the left chamber,
\begin{equation}
n_t=\sum_{i=1}^N b_i(t).
\label{eq:classical_macrostate}
\end{equation}
The Boltzmann entropy of this macrostate is
\begin{equation}
S_B(n_t)=\log {\binom{N}{n_t}}.
\label{eq:classical_entropy}
\end{equation}
At each time step, a particle switches chamber with probability \(\alpha\) and remains in the same chamber with probability \(\beta=1-\alpha\). The corresponding \(k\)-step switching and staying probabilities are
\begin{equation}
\alpha_k=\frac{1}{2}\left(1-\gamma^k\right),
\qquad
\beta_k=\frac{1}{2}\left(1+\gamma^k\right),
\qquad
\gamma=\beta-\alpha .
\label{eq:alpha_beta_k}
\end{equation}

Let \(T_k(m,n)\) denote the probability that the system evolves from \(n\) particles in the left chamber to \(m\) particles in the left chamber after \(k\) steps. This transition probability is
\begin{equation}
T_k(m,n)
=
\sum_{i=\max(0,n-m)}^{\min(n,N-m)}
{n\choose i}
\alpha_k^i
\beta_k^{n-i}
{N-n\choose i+m-n}
\alpha_k^{i+m-n}
\beta_k^{N-m-i}.
\label{eq:classical_transition_kernel}
\end{equation}
Here \(i\) counts the number of particles that move from left to right, while \(i+m-n\) counts the number of particles that move from right to left. The sum includes all microscopic transition counts compatible with the macroscopic change from \(n\) to \(m\).

Now impose the two endpoint conditions
\begin{equation}
a:n_0=0,
\qquad
b:n_{t_f}=n_f .
\label{eq:classical_endpoint_conditions}
\end{equation}
The first condition is the Past Hypothesis in this model: initially, all particles are in one chamber, so the macrostate has very low entropy. The second condition specifies the later macrostate. Using Eq.~\eqref{eq:classical_transition_kernel}, the bridge-conditioned macrostate distribution becomes
\begin{equation}
P(n_t=n\mid n_0=0,n_{t_f}=n_f)
=
\frac{
T_t(n,0)T_{t_f-t}(n_f,n)
}{
T_{t_f}(n_f,0)
}.
\label{eq:classical_bridge_distribution}
\end{equation}
Therefore, the classical bridge-conditioned entropy is
\begin{equation}
E[S_B(t)\mid n_0=0,n_{t_f}=n_f]
=
\sum_{n=0}^N
S_B(n)
\frac{
T_t(n,0)T_{t_f-t}(n_f,n)
}{
T_{t_f}(n_f,0)
}.
\label{eq:classical_bridge_entropy}
\end{equation}
By contrast, the past-only entropy profile is
\begin{equation}
E[S_B(t)\mid n_0=0]
=
\sum_{n=0}^N
S_B(n)T_t(n,0).
\label{eq:classical_past_only_entropy}
\end{equation}

The main result of the classical Boltzmann Bridge analysis is that the bridge-conditioned entropy in Eq.~\eqref{eq:classical_bridge_entropy} can behave very differently from the past-only entropy in Eq.~\eqref{eq:classical_past_only_entropy}. In particular, the bridge-conditioned entropy may rise above the entropy of the final macrostate and then decrease toward the final endpoint. To identify this behavior, one defines the sign diagnostic
\begin{equation}
\sigma(t,t_f)
=
\operatorname{Sign}
\left[
S_B(n_f)
-
E[S_B(t)\mid n_0=0,n_{t_f}=n_f]
\right].
\label{eq:classical_sign_diagnostic}
\end{equation}
When \(\sigma(t,t_f)=-1\), the expected entropy at the intermediate time is larger than the entropy of the final macrostate. This indicates that, among histories satisfying both endpoint conditions, the typical entropy profile decreases toward the final time. The point is not that the microscopic dynamics cease to be time-symmetric, nor that the usual relaxation curve is wrong. Rather, the point is that two-time conditioning selects a different ensemble of histories from the past-only ensemble.

\begin{figure}[h]
    \centering
    \includegraphics[width=0.9\linewidth]{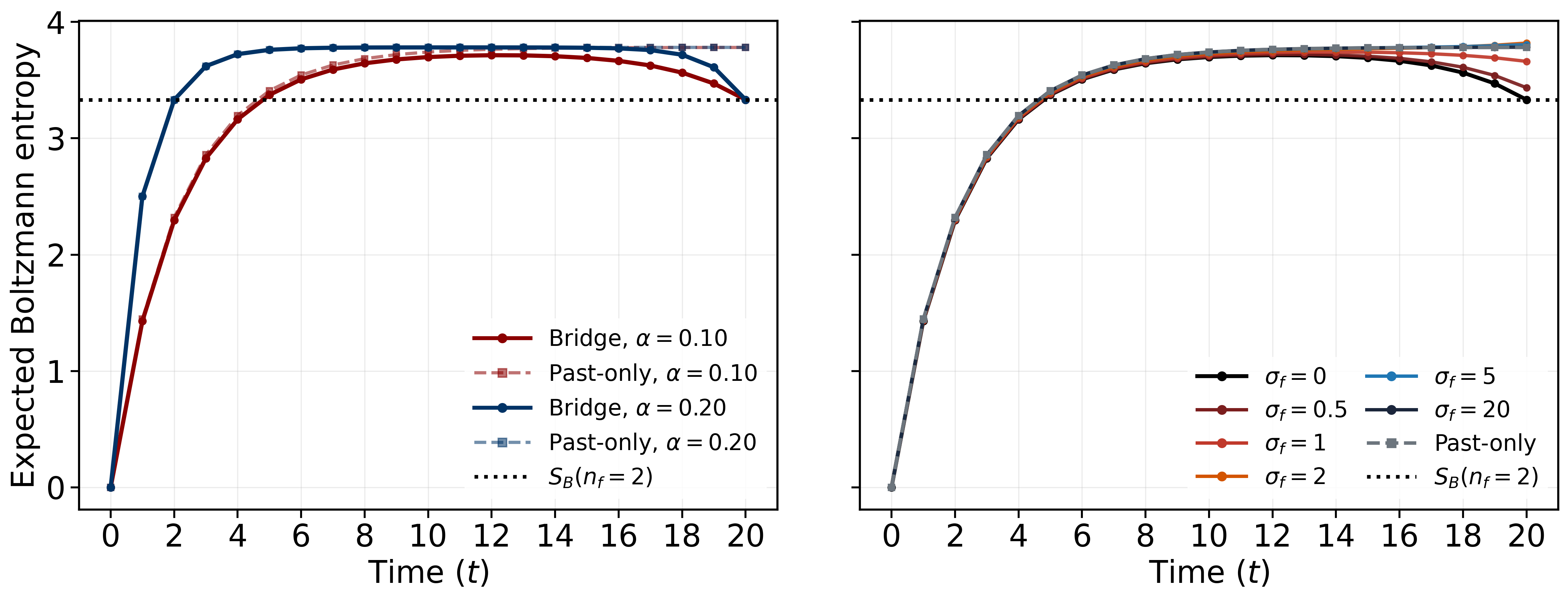}
    \caption{Comparison of bridge-conditioned and past-only entropy profiles. Left panel: Solid curves show the Boltzmann-bridge entropy profiles, while dashed curves show the corresponding entropy evolution conditioned only on the initial macrocondition, for two values of the switching probability $\alpha$. Right panel: Effect of relaxing the final macrocondition by replacing it with a noisy observation centered at $n_{f}$ with width $\sigma_{f}$. As $\sigma_f$ increases, the distinction between the bridge-conditioned and past-only entropy profiles becomes progressively smaller, since the future constraint becomes too noisy to provide useful conditioning information.}
    \label{fig:Classical_bridge}
\end{figure}

Figure~\ref{fig:Classical_bridge} illustrates the effect of two-time conditioning in the classical Boltzmann Bridge. In the left panel, we compare the bridge-conditioned entropy profile with the corresponding past-only entropy profile. The solid curves show the Boltzmann Bridge result for \(\alpha=0.1\) and \(\alpha=0.2\), while the dashed curves show the entropy evolution obtained when only the initial macrocondition \(n_0=0\) is imposed. The comparison shows that conditioning on the final macrostate changes the expected entropy history: the bridge curve is not simply the forward relaxation curve from the low-entropy initial state, but is reshaped by the later macrocondition.

In the right panel, we relax the exact final constraint by replacing the sharp condition \(n_{t_f}=n_f\) with a noisy final observation centered at \(n_f\). This is implemented using a Gaussian-like final weight with width \(\sigma_f\). As \(\sigma_f\) increases, the final macrocondition becomes less informative. Consequently, the bridge-conditioned entropy profile moves progressively toward the past-only curve. In the large-noise limit, the future constraint carries little useful information, and the relevant entropy history is therefore dominated by the initial macrocondition alone.

In the following sections, we use the classical Boltzmann Bridge as the reference model. Our goal is not to rederive the classical result, but to construct a quantum analogue of Eqs.~\eqref{eq:classical_bridge_distribution}--\eqref{eq:classical_bridge_entropy}. To do this, the classical macrostate \(n_t\) will be replaced by a quantum macro-observable, classical macrostates will be replaced by macro-subspace projectors, and stochastic transition probabilities will be replaced by unitary evolution combined with macroscopic conditioning. This leads to the quantum Boltzmann Bridge formulation developed next.

\section{Quantum Boltzmann Bridges}
\label{sec:quantum_boltzmann_bridges}

\subsection{General quantum bridge construction}
\label{subsec:quantum_bridge_construction}

The classical Boltzmann Bridge is formulated in terms of a stochastic process over macrostates. To extend this construction to quantum systems, we replace classical macrostates by macroscopic quantum subspaces and classical transition probabilities by unitary time evolution together with macroscopic coarse-graining. The goal is not to introduce dissipation or an external stochastic process by hand, but to define a bridge construction for a closed quantum system. The resulting object should be understood as a probability distribution over coarse-grained quantum macro-histories, rather than as an ordinary classical stochastic process \cite{AharonovBergmannLebowitz1964,Griffiths1984,Omnes1992,GellMannHartle1993}.

We consider a system of $N$ particles with full Hilbert space
\begin{equation}
\mathcal H
=
\bigotimes_{i=1}^{N}\mathcal H_i .
\label{eq:quantum_hilbert_space}
\end{equation}
The dynamics are assumed to be closed and unitary. Therefore, for a one-step unitary operator $U$, the density matrix evolves as
\begin{equation}
\rho_t
=
U^t\rho_0(U^\dagger)^t .
\label{eq:quantum_unitary_evolution}
\end{equation}
This choice keeps the quantum model as close as possible to microscopic reversible dynamics. Any apparent irreversibility or bridge-like behavior must therefore arise from coarse-graining and conditioning, rather than from an explicitly irreversible evolution equation.

The classical macroscopic variable in the two-chamber model is the number of particles in the left chamber. In the quantum setting, this becomes a macroscopic observable. Let $P_{L,i}$ denote the projector onto the left-chamber sector of particle $i$, understood as acting on the full tensor-product Hilbert space by identity on all other factors. We define
\begin{equation}
\hat n_L
=
\sum_{i=1}^{N}P_{L,i}.
\label{eq:quantum_macro_observable}
\end{equation}
The eigenvalues of this observable are
\begin{equation}
n=0,1,\ldots,N,
\label{eq:quantum_macro_eigenvalues}
\end{equation}
corresponding to the possible numbers of particles in the left chamber. Let $\Pi_n$ denote the projector onto the eigenspace of $\hat n_L$ with eigenvalue $n$. Thus,
\begin{equation}
\hat n_L\Pi_n
=
n\Pi_n .
\label{eq:quantum_macro_projector_definition}
\end{equation}
The projectors form an orthogonal resolution of the identity. They satisfy
\begin{equation}
\Pi_n\Pi_m
=
\delta_{nm}\Pi_n ,
\label{eq:quantum_projector_orthogonality}
\end{equation}
and
\begin{equation}
\sum_{n=0}^{N}\Pi_n
=
I .
\label{eq:quantum_projector_completeness}
\end{equation}
Thus, the quantum analogue of a classical macrostate is not a single state vector, but a macro-subspace of $\mathcal H$.

The entropy used in this work is the quantum Boltzmann entropy of a macrostate. It is defined by the dimension of the corresponding macro-subspace:
\begin{equation}
S_B(n)
=
\log \dim \operatorname{Ran}(\Pi_n).
\label{eq:quantum_boltzmann_entropy}
\end{equation}
Here $\operatorname{Ran}(\Pi_n)$ is the range of the projector $\Pi_n$. This definition is the direct quantum analogue of the classical Boltzmann entropy, where entropy is the logarithm of the number of microscopic configurations compatible with a given macrostate. In the quantum setting, the number of compatible microstates is replaced by the dimension of the Hilbert subspace compatible with the macroscopic condition \cite{PathriaBeale2011,GoldsteinLebowitzTumulkaZanghi2006,PopescuShortWinter2006}.

It is important to distinguish this entropy from the von Neumann entropy of the full quantum state,
\begin{equation}
S_{\mathrm{vN}}(\rho)
=
-\operatorname{Tr}(\rho\log\rho).
\label{eq:von_neumann_entropy}
\end{equation}
The von Neumann entropy is a property of the density matrix $\rho$, whereas the Boltzmann entropy in Eq.~\eqref{eq:quantum_boltzmann_entropy} is a property of a macro-subspace \cite{VonNeumann1927,Wehrl1978}. For the closed quantum systems considered here, the global von Neumann entropy is invariant under unitary time evolution:
\begin{equation}
S_{\mathrm{vN}}(\rho_t)
=
S_{\mathrm{vN}}(\rho_0).
\label{eq:vn_entropy_unitary_invariance}
\end{equation}
Therefore, the global von Neumann entropy cannot describe the bridge entropy profile studied here. In the minimal pure-state model, it would remain zero at all times. In the internally mixed model, it would remain fixed by the initial mixture over hidden internal states. In neither case would it measure how the system moves among macroscopic sectors labelled by $n$.

One could instead consider reduced or dephased entropies, such as the entropy of a subsystem or the Shannon entropy of the macrostate distribution. These quantities may be useful for related questions, but they answer a different question from the one studied here. The Boltzmann Bridge is concerned with the entropy assigned to macrostates and with the expected value of this macrostate entropy under one-time or two-time conditioning. For this reason, the appropriate quantum analogue in the present construction is the expected Boltzmann entropy over the macro-projector distribution, not the von Neumann entropy of the full state.

The quantum Past Hypothesis is represented by an initial state supported entirely in the low-entropy macro-subspace $n=0$. We write
\begin{equation}
\rho_0
=
\rho_a ,
\label{eq:quantum_initial_state}
\end{equation}
with
\begin{equation}
\rho_a
=
\Pi_0\rho_a\Pi_0 .
\label{eq:quantum_past_hypothesis}
\end{equation}
The final macrocondition is that, at time $t_f$, the system is found in the macro-sector with $n_f$ particles in the left chamber:
\begin{equation}
n_{t_f}
=
n_f .
\label{eq:quantum_final_condition}
\end{equation}
In quantum language, this final condition is represented by the projector $\Pi_{n_f}$. We denote the initial and final macroconditions by $a$ and $b$, respectively.

The past-only quantum macrostate distribution is obtained by applying the Born rule to the macro-projectors:
\begin{equation}
p_Q(n,t\mid a)
=
\operatorname{Tr}(\Pi_n\rho_t).
\label{eq:quantum_past_only_distribution}
\end{equation}
The corresponding past-only expected Boltzmann entropy is
\begin{equation}
E_Q[S_B(t)\mid a]
=
\sum_{n=0}^{N}
S_B(n)\,p_Q(n,t\mid a).
\label{eq:quantum_past_only_entropy}
\end{equation}
This is the quantum analogue of the classical relaxation curve conditioned only on the low-entropy initial macrocondition.

To define the quantum Boltzmann Bridge, we condition not only on the initial macrocondition but also on the final macrocondition. The bridge weight is obtained by projecting onto the intermediate macro-sector $n$, evolving the projected state to the final time, and then imposing the final macro-projector $\Pi_{n_f}$. The unnormalized bridge weight is
\begin{equation}
\widetilde p_Q(n,t;a,b)
=
\operatorname{Tr}
\left[
\Pi_{n_f}
U^{t_f-t}
\Pi_n\rho_t\Pi_n
(U^\dagger)^{t_f-t}
\right].
\label{eq:quantum_bridge_weight}
\end{equation}
Equivalently, Eq.~\eqref{eq:quantum_bridge_weight} is the joint probability for the coarse-grained macro-history in which the system is assigned to macro-sector $n$ at time $t$ and macro-sector $n_f$ at time $t_f$, under the initial macrocondition $a$.

The insertion of $\Pi_n\rho_t\Pi_n$ is essential. It represents a L\"uders projection, or equivalently a dephasing with respect to the intermediate macro-projector decomposition, before the remaining unitary evolution to the final time. Thus, the bridge probabilities defined here are not passive Bayesian conditionals of an unmeasured closed quantum trajectory. They are probabilities for coarse-grained macro-histories in which the intermediate macro-sector is specified. In regimes where the relevant macro-histories decohere, this construction coincides with the usual logic of consistent or decoherent histories; in general, it should be read as the operational macro-projective bridge construction used in this work \cite{Griffiths1984,Omnes1992,GellMannHartle1993}.

Normalizing over all possible intermediate macrostates gives
\begin{equation}
p_Q(n,t\mid a,b)
=
\frac{
\widetilde p_Q(n,t;a,b)
}{
\sum_{m=0}^{N}
\widetilde p_Q(m,t;a,b)
}.
\label{eq:quantum_bridge_distribution}
\end{equation}
This expression is defined whenever the denominator is nonzero. The bridge-conditioned expected Boltzmann entropy is then
\begin{equation}
E_Q[S_B(t)\mid a,b]
=
\sum_{n=0}^{N}
S_B(n)\,p_Q(n,t\mid a,b).
\label{eq:quantum_bridge_entropy}
\end{equation}

This definition mirrors the logic of the classical bridge while respecting the structure of quantum mechanics. The projectors $\Pi_n$ represent macroscopic coarse-graining, the unitary $U$ represents closed microscopic dynamics, and the final projector $\Pi_{n_f}$ implements the later macrocondition. However, because the intermediate macrocondition is implemented by projection, the denominator in Eq.~\eqref{eq:quantum_bridge_distribution} is the probability of the final macrocondition within the dephased macro-history ensemble. It is not, in general, identical to the uninterrupted final probability
\begin{equation}
\operatorname{Tr}
\left[
\Pi_{n_f}
U^{t_f}
\rho_a
(U^\dagger)^{t_f}
\right].
\label{eq:uninterrupted_final_probability}
\end{equation}
This distinction is one of the main differences between the classical bridge and its closed quantum analogue.

The quantum analogue of the classical sign diagnostic is
\begin{equation}
\sigma_Q(t,t_f)
=
\operatorname{Sign}
\left[
S_B(n_f)
-
E_Q[S_B(t)\mid a,b]
\right].
\label{eq:quantum_sign_diagnostic}
\end{equation}
When $\sigma_Q(t,t_f)=-1$, the bridge-conditioned expected Boltzmann entropy at the intermediate time is larger than the entropy of the final macrostate. This is the quantum analogue of the classical bridge regime in which the two-time-conditioned entropy profile bends downward toward the final endpoint.

The construction above is model-independent. It specifies how to define a quantum Boltzmann Bridge once the Hilbert space, macro-projectors, initial state, and unitary dynamics are chosen. In the next subsection, we study the simplest realization of this framework: the minimal coherent chamber-qubit model. In that model, each particle has only two quantum states, $|R\rangle$ and $|L\rangle$, and the one-particle dynamics are given by a unitary rotation. This model provides the direct quantum analogue of the classical left-right bit system and serves as the baseline quantum bridge, which we call the coherent chamber-qubit bridge.

\subsection{The coherent chamber-qubit bridge}
\label{subsec:coherent_chamber_qubit_bridge}

The first quantum realization of the Boltzmann Bridge is the minimal closed model in which each particle carries only a two-state chamber degree of freedom. We call this model the \emph{coherent chamber-qubit bridge}. It is the most direct quantum analogue of the classical left-right bit model: each classical bit is replaced by a qubit whose basis states represent the right and left chambers.

For each particle, the Hilbert space is
\begin{equation}
\mathcal H_i
=
\operatorname{span}\{|R\rangle_i,|L\rangle_i\}
\cong
\mathbb C^2 .
\label{eq:coherent_single_particle_hilbert}
\end{equation}
The full Hilbert space is therefore
\begin{equation}
\mathcal H
=
(\mathbb C^2)^{\otimes N}.
\label{eq:coherent_full_hilbert}
\end{equation}

The macroscopic observable is the number of particles in the left chamber. Let $P_{L,i}$ denote the projector onto the left-chamber state of particle $i$, with identity on all other tensor factors understood:
\begin{equation}
P_{L,i}
=
|L\rangle_i\langle L|_i .
\label{eq:coherent_left_projector}
\end{equation}
The macroscopic number operator is
\begin{equation}
\hat n_L
=
\sum_{i=1}^{N} P_{L,i}.
\label{eq:coherent_macro_observable}
\end{equation}

Let $\Pi_n$ denote the projector onto the subspace with exactly $n$ particles in the left chamber. Since this sector has dimension $\binom{N}{n}$, the quantum Boltzmann entropy is
\begin{equation}
S_B(n)
=
\log \binom{N}{n}.
\label{eq:coherent_boltzmann_entropy}
\end{equation}
Thus, in this minimal model, the quantum Boltzmann entropy exactly matches the classical two-box entropy.

The quantum Past Hypothesis is represented by the pure all-right state
\begin{equation}
\rho_a
=
|R\cdots R\rangle\langle R\cdots R|.
\label{eq:coherent_initial_density}
\end{equation}
Equivalently, the initial state is supported entirely in the $n=0$ macro-subspace. This is the quantum analogue of the classical condition that all particles begin in the right chamber.

The one-particle dynamics are chosen to be a unitary rotation in the chamber basis:
\begin{equation}
u(\theta)
=
\begin{pmatrix}
\cos\theta & -\sin\theta \\
\sin\theta & \cos\theta
\end{pmatrix},
\label{eq:coherent_one_particle_unitary}
\end{equation}
where the ordered basis is $(|R\rangle,|L\rangle)$. The full $N$-particle evolution is taken to be independent across particles:
\begin{equation}
U
=
u(\theta)^{\otimes N}.
\label{eq:coherent_many_particle_unitary}
\end{equation}
This choice preserves the closed and coherent nature of the quantum dynamics. No stochastic transition rule or external dissipation is imposed. Throughout this subsection, time is discrete. Thus \(t\), \(t_f\), and \(\Delta=t_f-t\) denote integer numbers of time steps.

For comparison with the classical model, we calibrate the one-step transition probability by setting
\begin{equation}
\sin^2\theta
=
\alpha .
\label{eq:coherent_alpha_calibration}
\end{equation}
With this calibration,
\begin{equation}
\cos^2\theta
=
1-\alpha
=
\beta .
\label{eq:coherent_beta_calibration}
\end{equation}
Thus, the one-step quantum probability of moving from $|R\rangle$ to $|L\rangle$ agrees with the classical one-step switching probability. However, this agreement holds only at the one-step level. The multi-step quantum dynamics are unitary and coherent, not Markovian.

Since
\begin{equation}
u(\theta)^t
=
u(t\theta),
\label{eq:coherent_rotation_power}
\end{equation}
a single particle initially in the right chamber evolves as
\begin{equation}
u(\theta)^t |R\rangle
=
\cos(t\theta)|R\rangle
+
\sin(t\theta)|L\rangle .
\label{eq:coherent_single_particle_evolution}
\end{equation}
For convenience, define
\begin{equation}
c_t
=
\cos(t\theta),
\qquad
s_t
=
\sin(t\theta).
\label{eq:coherent_ct_st_def}
\end{equation}
Then the $N$-particle state at time $t$ is
\begin{equation}
|\psi_t\rangle
=
(c_t|R\rangle+s_t|L\rangle)^{\otimes N}.
\label{eq:coherent_many_particle_state}
\end{equation}

Expanding this state in the symmetric Dicke basis gives
\begin{equation}
|\psi_t\rangle
=
\sum_{n=0}^{N}
\sqrt{\binom{N}{n}}\,
c_t^{N-n}
s_t^{n}
|D_n\rangle ,
\label{eq:coherent_dicke_expansion}
\end{equation}
where $|D_n\rangle$ is the normalized symmetric state with exactly $n$ particles in the left chamber.

The past-only macrostate distribution follows directly from the Born rule:
\begin{equation}
p_Q^{\mathrm{coh}}(n,t\mid a)
=
\operatorname{Tr}(\Pi_n\rho_t)
=
\binom{N}{n}
c_t^{2(N-n)}
s_t^{2n}.
\label{eq:coherent_past_only_distribution}
\end{equation}
The corresponding past-only expected Boltzmann entropy is
\begin{equation}
E_Q^{\mathrm{coh}}[S_B(t)\mid a]
=
\sum_{n=0}^{N}
S_B(n)
\binom{N}{n}
c_t^{2(N-n)}
s_t^{2n}.
\label{eq:coherent_past_only_entropy}
\end{equation}

This has the same binomial form as the classical past-only distribution, but with an important difference: the effective one-particle left probability is
\begin{equation}
p_t
=
s_t^2
=
\sin^2(t\theta).
\label{eq:coherent_effective_probability}
\end{equation}
This probability is oscillatory. In contrast, the classical stochastic probability relaxes toward \(1/2\). More explicitly, the coherent model gives the multi-step left probability \(\sin^2(t\theta)\), whereas the classical Markov model gives the \(t\)-step switching probability \(\alpha_t=\frac{1}{2}(1-\gamma^t)\). These agree at one step by construction, but not at later times. Therefore, even before imposing the final bridge condition, the coherent chamber-qubit model does not describe classical relaxation.

We now impose a final macrocondition
\begin{equation}
b:\ n_{t_f}=n_f .
\label{eq:coherent_final_condition}
\end{equation}
Let
\begin{equation}
\Delta
=
t_f-t
\label{eq:coherent_delta_def}
\end{equation}
be the remaining number of time steps between the intermediate macro-conditioning and the final macro-conditioning. The unnormalized bridge weight is
\begin{equation}
\widetilde p_Q^{\mathrm{coh}}(n,t;a,b)
=
\operatorname{Tr}
\left[
\Pi_{n_f}
U^\Delta
\Pi_n\rho_t\Pi_n
(U^\dagger)^\Delta
\right].
\label{eq:coherent_bridge_weight_general}
\end{equation}

Because the initial state is permutation-symmetric and $U=u(\theta)^{\otimes N}$ preserves the symmetric subspace, the projected state $\Pi_n\rho_t\Pi_n$ is proportional to $|D_n\rangle\langle D_n|$. Explicitly,
\begin{equation}
\Pi_n\rho_t\Pi_n
=
\binom{N}{n}
c_t^{2(N-n)}
s_t^{2n}
|D_n\rangle\langle D_n|.
\label{eq:coherent_projected_density}
\end{equation}
Define the Dicke-sector transition amplitude
\begin{equation}
A_{m,n}(\Delta)
=
\langle D_m|U^\Delta|D_n\rangle .
\label{eq:coherent_dicke_amplitude}
\end{equation}
Then the unnormalized bridge weight becomes
\begin{equation}
\widetilde p_Q^{\mathrm{coh}}(n,t;a,b)
=
\binom{N}{n}
c_t^{2(N-n)}
s_t^{2n}
\left|A_{n_f,n}(\Delta)\right|^2 .
\label{eq:coherent_bridge_weight}
\end{equation}
For a detailed derivation of the Dicke-state amplitudes used in the coherent bridge model, see Appendix~ \ref{app:dicke_amplitude_coherent_bridge}.

Normalizing over all intermediate macrostates gives the bridge-conditioned distribution
\begin{equation}
p_Q^{\mathrm{coh}}(n,t\mid a,b)
=
\frac{
\binom{N}{n}
c_t^{2(N-n)}
s_t^{2n}
\left|A_{n_f,n}(\Delta)\right|^2
}{
\sum_{m=0}^{N}
\binom{N}{m}
c_t^{2(N-m)}
s_t^{2m}
\left|A_{n_f,m}(\Delta)\right|^2
}.
\label{eq:coherent_bridge_distribution}
\end{equation}
This expression is defined whenever the denominator is nonzero. The bridge-conditioned expected entropy is therefore
\begin{equation}
E_Q^{\mathrm{coh}}[S_B(t)\mid a,b]
=
\sum_{n=0}^{N}
S_B(n)\,
p_Q^{\mathrm{coh}}(n,t\mid a,b).
\label{eq:coherent_bridge_entropy}
\end{equation}
The corresponding sign diagnostic is
\begin{equation}
\sigma_Q^{\mathrm{coh}}(t,t_f)
=
\operatorname{Sign}
\left[
S_B(n_f)
-
E_Q^{\mathrm{coh}}[S_B(t)\mid a,b]
\right].
\label{eq:coherent_sign_diagnostic}
\end{equation}
A negative value of $\sigma_Q^{\mathrm{coh}}$ means that the bridge-conditioned expected entropy at time $t$ is larger than the entropy of the final macrostate.

The coherent chamber-qubit bridge is useful precisely because it is minimal. It preserves the same macroscopic observable and the same Boltzmann entropy as the classical two-box model, while replacing the classical Markov process by closed unitary dynamics. However, this minimality also limits its ability to reproduce classical behavior. Since the evolution is fully coherent, the macrostate distribution can undergo revivals and oscillations. As a result, negative values of $\sigma_Q^{\mathrm{coh}}$, when they occur, should not automatically be interpreted as classical bridge relaxation. They may instead arise from coherent quantum recurrence.

For this reason, the coherent chamber-qubit bridge should be viewed as the baseline quantum model rather than as the expected classical limit. It answers the question of what happens if the classical left-right bit is quantized as directly as possible. In the next subsection, we introduce a richer closed quantum model in which each chamber sector contains hidden internal microstates. This hidden-microstate bridge remains unitary at the microscopic level, but the unresolved internal degrees of freedom can suppress simple coherent revivals and allow the coarse-grained bridge behavior to become more classical-looking.

\subsection{The hidden-microstate bridge}
\label{subsec:hidden_microstate_bridge}

The coherent chamber-qubit bridge is the most direct closed quantum analogue of the classical left-right bit model. However, its Hilbert space is too small to contain any unresolved internal structure. Each particle only carries a chamber label, and the resulting dynamics remain strongly coherent and revival-dominated. To obtain a more thermodynamic quantum model, we now enrich each chamber sector with hidden internal microstates. This is consistent with the broader idea that coarse macroscopic behavior in closed quantum systems can depend strongly on unresolved internal degrees of freedom and typicality within large Hilbert spaces \cite{GoldsteinLebowitzTumulkaZanghi2006,PopescuShortWinter2006}.

We call this model the \emph{hidden-microstate bridge}. For each particle, the Hilbert space is decomposed as
\begin{equation}
\mathcal H_i
=
\mathcal H_{R,i}\oplus \mathcal H_{L,i},
\qquad
\dim \mathcal H_{R,i}
=
\dim \mathcal H_{L,i}
=
M .
\label{eq:hidden_single_particle_hilbert}
\end{equation}
Here \(M\) is the number of unresolved internal microstates inside each chamber sector. We choose orthonormal bases
\begin{equation}
\{|R,\mu\rangle_i\}_{\mu=1}^{M},
\qquad
\{|L,\mu\rangle_i\}_{\mu=1}^{M}.
\label{eq:hidden_internal_basis}
\end{equation}
Thus, each particle carries both a coarse chamber label and an internal label. The full Hilbert space is
\begin{equation}
\mathcal H
=
\bigotimes_{i=1}^{N}\mathcal H_i .
\label{eq:hidden_full_hilbert}
\end{equation}

For the one-particle Hilbert space, define the right- and left-chamber projectors by
\begin{equation}
P_R
=
\sum_{\mu=1}^{M}
|R,\mu\rangle\langle R,\mu|,
\qquad
P_L
=
\sum_{\mu=1}^{M}
|L,\mu\rangle\langle L,\mu|.
\label{eq:hidden_one_particle_projectors}
\end{equation}
Equivalently, \(P_R\) and \(P_L\) are the identities on the right- and left-chamber internal subspaces. For particle \(i\), the corresponding left-chamber projector acting on the full tensor-product Hilbert space is denoted by \(P_{L,i}\), with identity on all other tensor factors understood:
\begin{equation}
P_{L,i}
=
\sum_{\mu=1}^{M}
|L,\mu\rangle_i\langle L,\mu|_i .
\label{eq:hidden_left_projector}
\end{equation}
The macroscopic observable is again the number of particles in the left chamber:
\begin{equation}
\hat n_L
=
\sum_{i=1}^{N}P_{L,i}.
\label{eq:hidden_macro_observable}
\end{equation}
Let \(\Pi_n\) denote the projector onto the eigenspace of \(\hat n_L\) with eigenvalue \(n\). This is the macro-subspace with exactly \(n\) particles in the left chamber.

The dimension of this macro-subspace is now
\begin{equation}
\dim \operatorname{Ran}(\Pi_n)
=
\binom{N}{n}M^N .
\label{eq:hidden_macro_dimension}
\end{equation}
Therefore, the quantum Boltzmann entropy is
\begin{equation}
S_B^{(M)}(n)
=
\log\left(\binom{N}{n}M^N\right)
=
\log \binom{N}{n}
+
N\log M .
\label{eq:hidden_boltzmann_entropy}
\end{equation}
The additive term \(N\log M\) is independent of \(n\). Hence, when comparing bridge shapes or sign diagnostics, we may equivalently use the reduced entropy
\begin{equation}
S_{\mathrm{red}}(n)
=
\log \binom{N}{n}.
\label{eq:hidden_reduced_entropy}
\end{equation}
This reduced entropy allows direct comparison with both the classical bridge and the coherent chamber-qubit bridge.

The quantum Past Hypothesis is represented by a state that is macroscopically all-right but microscopically unresolved. We take the initial state to be maximally mixed over the all-right internal subspace:
\begin{equation}
\rho_a
=
\left(\frac{I_R}{M}\right)^{\otimes N},
\qquad
I_R
=
\sum_{\mu=1}^{M}
|R,\mu\rangle\langle R,\mu|.
\label{eq:hidden_initial_state}
\end{equation}
Equivalently,
\begin{equation}
\rho_a
=
\frac{\Pi_0}{\operatorname{Tr}(\Pi_0)} .
\label{eq:hidden_initial_projector_form}
\end{equation}
This expresses the fact that the macro-condition \(n_0=0\) is known, while the internal microstate within the all-right macro-subspace is not known.

The one-particle unitary acts on \(\mathcal H_R\oplus \mathcal H_L\). In block form,
\begin{equation}
u
=
\begin{pmatrix}
A & B\\
C & D
\end{pmatrix},
\label{eq:hidden_general_block_unitary}
\end{equation}
where
\begin{equation}
A:\mathcal H_R\to\mathcal H_R,
\qquad
B:\mathcal H_L\to\mathcal H_R,
\qquad
C:\mathcal H_R\to\mathcal H_L,
\qquad
D:\mathcal H_L\to\mathcal H_L .
\label{eq:hidden_block_meaning}
\end{equation}
The full \(N\)-particle dynamics are again taken to factorize:
\begin{equation}
U
=
u^{\otimes N}.
\label{eq:hidden_many_particle_unitary}
\end{equation}
Throughout this subsection, time is discrete. Thus \(t\), \(t_f\), and \(\Delta=t_f-t\) denote integer numbers of time steps.

For the numerical and comparative analysis, we use the structured one-particle family
\begin{equation}
u
=
\begin{pmatrix}
\cos\theta\,U_R & -\sin\theta\,U_R\\
\sin\theta\,U_L & \cos\theta\,U_L
\end{pmatrix},
\qquad
U_R,U_L\in U(M).
\label{eq:hidden_structured_unitary}
\end{equation}
This matrix is unitary because it is the product of a chamber rotation and internal unitaries \(U_R\) and \(U_L\). The chamber-transition scale is controlled by \(\theta\), while \(U_R\) and \(U_L\) generate internal unitary dynamics inside the right and left chamber sectors. To compare with the classical switching probability \(\alpha\), we use the calibration
\begin{equation}
\sin^2\theta
=
\alpha .
\label{eq:hidden_theta_calibration}
\end{equation}
This matches the one-step chamber-switching probability to the classical value, while still allowing the multi-step quantum dynamics to differ from the classical Markov process.

Let the \(t\)-step one-particle unitary be written as
\begin{equation}
u^t
=
\begin{pmatrix}
A_t & B_t\\
C_t & D_t
\end{pmatrix}.
\label{eq:hidden_t_step_unitary}
\end{equation}
For a single particle initially in the right-sector microcanonical state \(I_R/M\), the time-\(t\) state is
\begin{equation}
\rho_t^{(1)}
=
u^t\frac{I_R}{M}(u^\dagger)^t .
\label{eq:hidden_one_particle_state}
\end{equation}
The coarse probability that the particle is in the left chamber at time \(t\) is
\begin{equation}
p_t
=
\operatorname{Tr}\left(P_L\rho_t^{(1)}\right)
=
\frac{1}{M}\operatorname{Tr}\left(C_t^\dagger C_t\right).
\label{eq:hidden_left_probability}
\end{equation}
Similarly,
\begin{equation}
1-p_t
=
\operatorname{Tr}\left(P_R\rho_t^{(1)}\right)
=
\frac{1}{M}\operatorname{Tr}\left(A_t^\dagger A_t\right).
\label{eq:hidden_right_probability}
\end{equation}

Because the initial state is a product state and the unitary evolution factorizes across particles, the past-only macrostate distribution remains binomial at the coarse level:
\begin{equation}
p_Q^{(M)}(n,t\mid a)
=
\binom{N}{n}p_t^n(1-p_t)^{N-n}.
\label{eq:hidden_past_only_distribution}
\end{equation}
The corresponding past-only expected entropy is
\begin{equation}
E_Q^{(M)}[S_B(t)\mid a]
=
\sum_{n=0}^{N}
S_B^{(M)}(n)
\binom{N}{n}p_t^n(1-p_t)^{N-n}.
\label{eq:hidden_past_only_entropy}
\end{equation}
For comparisons of bridge shape and sign, we use the reduced form
\begin{equation}
E_{Q,\mathrm{red}}^{(M)}[S_{\mathrm{red}}(t)\mid a]
=
\sum_{n=0}^{N}
S_{\mathrm{red}}(n)
\binom{N}{n}p_t^n(1-p_t)^{N-n}.
\label{eq:hidden_past_only_entropy_reduced}
\end{equation}

We now impose the final macrocondition
\begin{equation}
b:\ n_{t_f}=n_f,
\label{eq:hidden_final_condition}
\end{equation}
and define the remaining time interval
\begin{equation}
\Delta
=
t_f-t .
\label{eq:hidden_delta}
\end{equation}
The exact unnormalized bridge weight is
\begin{equation}
\widetilde p_Q^{(M)}(n,t;a,b)
=
\operatorname{Tr}
\left[
\Pi_{n_f}
U^\Delta
\Pi_n\rho_t\Pi_n
(U^\dagger)^\Delta
\right].
\label{eq:hidden_bridge_weight}
\end{equation}
To isolate the role of the intermediate macro-sector, define the normalized post-measurement state, for sectors with \(p_Q^{(M)}(n,t\mid a)>0\), by
\begin{equation}
\omega_n(t)
=
\frac{\Pi_n\rho_t\Pi_n}{p_Q^{(M)}(n,t\mid a)} .
\label{eq:hidden_omega_n}
\end{equation}
Then the bridge weight factorizes as
\begin{equation}
\widetilde p_Q^{(M)}(n,t;a,b)
=
p_Q^{(M)}(n,t\mid a)\,
K_{n_f,n}^{(M)}(t,\Delta),
\label{eq:hidden_bridge_weight_factorized}
\end{equation}
where the exact coarse propagator is
\begin{equation}
K_{m,n}^{(M)}(t,\Delta)
=
\operatorname{Tr}
\left[
\Pi_m
U^\Delta
\omega_n(t)
(U^\dagger)^\Delta
\right].
\label{eq:hidden_exact_K}
\end{equation}
For each fixed intermediate sector \(n\), this propagator is nonnegative and normalized over the final macro-sector:
\begin{equation}
K_{m,n}^{(M)}(t,\Delta)
\geq
0,
\qquad
\sum_{m=0}^{N}
K_{m,n}^{(M)}(t,\Delta)
=
1.
\label{eq:hidden_K_properties}
\end{equation}
The bridge-conditioned distribution is therefore
\begin{equation}
p_Q^{(M)}(n,t\mid a,b)
=
\frac{
p_Q^{(M)}(n,t\mid a)
K_{n_f,n}^{(M)}(t,\Delta)
}{
\sum_{m=0}^{N}
p_Q^{(M)}(m,t\mid a)
K_{n_f,m}^{(M)}(t,\Delta)
}.
\label{eq:hidden_bridge_distribution}
\end{equation}
This expression is defined whenever the denominator is nonzero. The detailed calculation of the hidden-microstate transition kernel ($K$) is provided in Appendix~\ref{app:hidden_microstate_K}. The bridge-conditioned entropy profile is
\begin{equation}
E_Q^{(M)}[S_B(t)\mid a,b]
=
\sum_{n=0}^{N}
S_B^{(M)}(n)\,
p_Q^{(M)}(n,t\mid a,b).
\label{eq:hidden_bridge_entropy}
\end{equation}
Equivalently, for comparison with the classical and coherent chamber-qubit bridges, we use
\begin{equation}
E_{Q,\mathrm{red}}^{(M)}[S_{\mathrm{red}}(t)\mid a,b]
=
\sum_{n=0}^{N}
S_{\mathrm{red}}(n)\,
p_Q^{(M)}(n,t\mid a,b).
\label{eq:hidden_bridge_entropy_reduced}
\end{equation}

The corresponding sign diagnostic is
\begin{equation}
\sigma_Q^{(M)}(t,t_f)
=
\operatorname{Sign}
\left[
S_{\mathrm{red}}(n_f)
-
E_{Q,\mathrm{red}}^{(M)}[S_{\mathrm{red}}(t)\mid a,b]
\right].
\label{eq:hidden_sign_diagnostic}
\end{equation}
A negative value of \(\sigma_Q^{(M)}\) means that the bridge-conditioned expected reduced entropy at the intermediate time is larger than the entropy of the final macrostate.

The central new object in the hidden-microstate bridge is the propagator \(K_{m,n}^{(M)}(t,\Delta)\). Unlike the Dicke amplitude in the coherent chamber-qubit bridge, this propagator depends not only on the remaining time \(\Delta\), but also on the internal state \(\omega_n(t)\) generated by the earlier evolution. This is the key source of richer behavior. The hidden internal degrees of freedom can scramble phases inside the coarse chamber sectors while the total dynamics remain closed and unitary.

In regimes where interference between coarse chamber histories becomes small, the exact propagator \(K_{m,n}^{(M)}(t,\Delta)\) is expected to behave like a classical-looking coarse transition kernel. This is the mechanism by which the hidden-microstate bridge can become more similar to the classical Boltzmann Bridge as \(M\) increases. The numerical comparison over different values of \(M\) will be given in the next subsection. There, we will show that small internal spaces remain strongly sample-dependent, while larger hidden internal spaces statistically recover the classical bridge sign structure.

\subsection{Numerical comparison of classical, coherent, and hidden-microstate bridges}
\label{subsec:numerical_comparison}

We now compare the three bridge constructions under the same benchmark conditions:
\begin{equation}
N=8,
\qquad
\alpha=0.1,
\qquad
\beta=0.9,
\qquad
t_f=20,
\qquad
n_f=2.
\label{eq:comparison_benchmark}
\end{equation}
The final reduced Boltzmann entropy is
\begin{equation}
S_{\mathrm{red}}(n_f)
=
S_{\mathrm{red}}(2)
=
\log \binom{8}{2}
=
\log(28)
=
3.332205.
\label{eq:comparison_final_entropy}
\end{equation}
For all models, we use the same reduced-entropy sign diagnostic:
\begin{equation}
\sigma(t,t_f)
=
\operatorname{Sign}
\left[
S_{\mathrm{red}}(n_f)
-
E_{\mathrm{red}}(t\mid a,b)
\right].
\label{eq:comparison_sign}
\end{equation}
Thus, \(\sigma=-1\) means that the bridge-conditioned expected reduced entropy at the intermediate time is larger than the entropy of the final macrostate. This is the bridge regime in which the expected entropy bends downward toward the final endpoint.

\subsubsection{One-to-one comparison}

Table~\ref{tab:comparison_three_models} compares the classical bridge, the coherent chamber-qubit bridge, and one concrete \(M=2\) hidden-microstate bridge. The same values of \(N\), \(\alpha\), \(t_f\), and \(n_f\) are used in all cases. The \(M=2\) hidden-microstate entry corresponds to one fixed realization of the internal unitaries and is therefore illustrative rather than ensemble-averaged.

\begin{table}[t]
\centering
\begin{tabular}{c c c c c c c c}
\hline
\(t\) & \(\Delta\)
& \(E_{\mathrm{cl}}\) & \(\sigma_{\mathrm{cl}}\)
& \(E_{\mathrm{coh}}\) & \(\sigma_{\mathrm{coh}}\)
& \(E_{M=2}\) & \(\sigma_{M=2}\) \\
\hline
2  & 18 & 2.294973 & \(+1\) & 3.374524 & \(-1\) & 3.121726 & \(+1\) \\
4  & 16 & 3.162978 & \(+1\) & 0.626360 & \(+1\) & 3.612803 & \(-1\) \\
6  & 14 & 3.506918 & \(-1\) & 2.349955 & \(+1\) & 3.723476 & \(-1\) \\
8  & 12 & 3.644939 & \(-1\) & 2.744718 & \(+1\) & 1.366661 & \(+1\) \\
10 & 10 & 3.698625 & \(-1\) & 1.906471 & \(+1\) & 1.706580 & \(+1\) \\
12 & 8  & 3.714362 & \(-1\) & 3.613969 & \(-1\) & 3.138269 & \(+1\) \\
14 & 6  & 3.705754 & \(-1\) & 0.649413 & \(+1\) & 3.734404 & \(-1\) \\
16 & 4  & 3.666051 & \(-1\) & 2.273863 & \(+1\) & 3.540005 & \(-1\) \\
18 & 2  & 3.564669 & \(-1\) & 2.116221 & \(+1\) & 0.632958 & \(+1\) \\
\hline
\end{tabular}
\caption{
One-to-one comparison of the classical bridge, the coherent chamber-qubit bridge, and one concrete \(M=2\) hidden-microstate bridge for \(N=8\), \(\alpha=0.1\), \(t_f=20\), and \(n_f=2\). The entropy values are reduced Boltzmann entropies, so they can be compared directly across the three models.
}
\label{tab:comparison_three_models}
\end{table}

The classical bridge gives the sign sequence
\begin{equation}
\sigma_{\mathrm{cl}}(t=2,4,6,8,10,12,14,16,18)
=
(+,+,-,-,-,-,-,-,-).
\label{eq:classical_sign_sequence}
\end{equation}
Thus, from \(t=6\) onward, the classical bridge enters a stable negative-sign region. This means that the expected bridge entropy is larger than the final entropy over most of the interval and then decreases toward the final macrocondition.

The coherent chamber-qubit bridge gives a very different sequence:
\begin{equation}
\sigma_{\mathrm{coh}}(t=2,4,6,8,10,12,14,16,18)
=
(-,+,+,+,+,-,+,+,+).
\label{eq:coherent_sign_sequence}
\end{equation}
This model can produce negative signs, but they appear at isolated times rather than forming a stable classical-like interval. This confirms that the coherent chamber-qubit bridge is not a classical relaxation model. Its behavior is dominated by coherent oscillations and revivals.

The concrete \(M=2\) hidden-microstate bridge gives
\begin{equation}
\sigma_{M=2}(t=2,4,6,8,10,12,14,16,18)
=
(+,-,-,+,+,+,-,-,+).
\label{eq:M2_sign_sequence}
\end{equation}
This already differs from the coherent chamber-qubit bridge, showing that hidden internal microstates reshape the bridge profile. However, \(M=2\) is still too small to produce a robust classical-like sign pattern. It produces scattered negative regions rather than a stable negative interval matching the classical bridge.

\subsubsection{Systematic dependence on hidden internal dimension}

The previous comparison used one concrete \(M=2\) hidden-microstate realization. To test whether the hidden-microstate bridge statistically approaches the classical bridge as the internal Hilbert space becomes larger, we now compare the internal dimensions
\begin{equation}
M=2,4,8,16,32.
\label{eq:M_values}
\end{equation}
For each value of \(M\), independent internal unitaries \(U_R,U_L\in U(M)\) are sampled from the Haar measure \cite{Mezzadri2007}. Since the bridge sign depends on the sampled internal unitaries, the relevant ensemble quantity is the negative fraction
\begin{equation}
f_-(t)
=
\Pr_{U_R,U_L}
\left[
\sigma_Q^{(M)}(t)=-1
\right].
\label{eq:negative_fraction}
\end{equation}
Here \(f_-(t)=1\) means that all sampled internal unitaries produce a negative bridge sign, while \(f_-(t)=0\) means that none do. Intermediate values indicate mixed behavior across the random-unitary ensemble. In the numerical results below, \(f_-(t)\) is estimated from 200 independent samples for each value of \(M\) and each time \(t\). The corresponding binomial sampling uncertainty is of order \(\sqrt{f_-(t)(1-f_-(t))/200}\).

\begin{table}[H]
\centering
\begin{tabular}{c c c c c c}
\hline
\(t\) & \(M=2\) & \(M=4\) & \(M=8\) & \(M=16\) & \(M=32\) \\
\hline
2  & 0.010 & 0.000 & 0.000 & 0.000 & 0.000 \\
4  & 0.230 & 0.180 & 0.105 & 0.060 & 0.005 \\
6  & 0.260 & 0.410 & 0.550 & 0.830 & 0.980 \\
8  & 0.120 & 0.435 & 0.725 & 0.965 & 1.000 \\
10 & 0.100 & 0.395 & 0.785 & 0.990 & 1.000 \\
12 & 0.125 & 0.300 & 0.735 & 1.000 & 1.000 \\
14 & 0.190 & 0.215 & 0.575 & 0.990 & 1.000 \\
16 & 0.205 & 0.165 & 0.395 & 0.980 & 1.000 \\
18 & 0.220 & 0.085 & 0.195 & 0.865 & 1.000 \\
\hline
\end{tabular}
\caption{
Negative fraction \(f_-(t)\) for the hidden-microstate bridge over internal dimensions \(M=2,4,8,16,32\). The benchmark is \(N=8\), \(\alpha=0.1\), \(t_f=20\), and \(n_f=2\). Each entry is estimated from 200 independent samples of \(U_R\) and \(U_L\).
}
\label{tab:negative_fraction_M}
\end{table}

Table~\ref{tab:negative_fraction_M} shows a clear trend. For \(M=2\) and \(M=4\), the signs are strongly sample-dependent and mostly positive. These internal spaces are too small to produce a robust classical-like bridge. For \(M=8\), the negative fraction exceeds \(1/2\) in the middle of the interval, especially for \(t=6,8,10,12,14\). This marks the onset of a classical-like negative bridge region. For \(M=16\), the negative fraction is close to one throughout the classical negative region. For \(M=32\), the hidden-microstate bridge nearly reproduces the classical sign pattern at the ensemble level.

\begin{table}[h]
\centering
\begin{tabular}{c c c c c c c c c c}
\hline
 & \(t=2\) & \(4\) & \(6\) & \(8\) & \(10\) & \(12\) & \(14\) & \(16\) & \(18\) \\
\hline
Classical & \(+\) & \(+\) & \(-\) & \(-\) & \(-\) & \(-\) & \(-\) & \(-\) & \(-\) \\
\(M=2\)  & \(+\mathrm{mix}\) & \(+\mathrm{mix}\) & \(+\mathrm{mix}\) & \(+\mathrm{mix}\) & \(+\mathrm{mix}\) & \(+\mathrm{mix}\) & \(+\mathrm{mix}\) & \(+\mathrm{mix}\) & \(+\mathrm{mix}\) \\
\(M=4\)  & \(+\) & \(+\mathrm{mix}\) & \(+\mathrm{mix}\) & \(+\mathrm{mix}\) & \(+\mathrm{mix}\) & \(+\mathrm{mix}\) & \(+\mathrm{mix}\) & \(+\mathrm{mix}\) & \(+\mathrm{mix}\) \\
\(M=8\)  & \(+\) & \(+\mathrm{mix}\) & \(-\mathrm{mix}\) & \(-\mathrm{mix}\) & \(-\mathrm{mix}\) & \(-\mathrm{mix}\) & \(-\mathrm{mix}\) & \(+\mathrm{mix}\) & \(+\mathrm{mix}\) \\
\(M=16\) & \(+\) & \(+\mathrm{mix}\) & \(-\mathrm{mix}\) & \(-\mathrm{mix}\) & \(-\mathrm{mix}\) & \(-\) & \(-\mathrm{mix}\) & \(-\mathrm{mix}\) & \(-\mathrm{mix}\) \\
\(M=32\) & \(+\) & \(+\mathrm{mix}\) & \(-\mathrm{mix}\) & \(-\) & \(-\) & \(-\) & \(-\) & \(-\) & \(-\) \\
\hline
\end{tabular}
\caption{
Qualitative sign-pattern comparison for the classical bridge and hidden-microstate bridges with increasing internal dimension. The labels \(+\mathrm{mix}\) and \(-\mathrm{mix}\) indicate mixed ensembles with positive-majority and negative-majority signs, respectively.
}
\label{tab:qualitative_sign_pattern_M}
\end{table}

This trend can be summarized by the qualitative sign-pattern table below. The symbols in Table~\ref{tab:qualitative_sign_pattern_M} are determined from \(f_-(t)\). A pure \(+\) means that \(f_-(t)=0\), so all sampled realizations have positive sign. A pure \(-\) means that \(f_-(t)=1\), so all sampled realizations have negative sign. The symbol \(+\mathrm{mix}\) means \(0<f_-(t)<1/2\): the ensemble is mixed, but positive signs are more common than negative signs. The symbol \(-\mathrm{mix}\) means \(1/2<f_-(t)<1\): the ensemble is mixed, but negative signs are more common than positive signs. Thus, the ``mix'' label indicates sample dependence, while the sign in front of ``mix'' indicates the majority sign.

The comparison shows that the classical bridge is not recovered by simply quantizing the chamber variable. The coherent chamber-qubit bridge remains dominated by unitary revivals. A small hidden internal space, such as \(M=2\), can modify the sign structure but still does not yield a stable classical-like bridge. As \(M\) increases, the hidden-microstate bridge statistically approaches the classical bridge sign pattern. This supports the interpretation that unresolved internal structure can suppress simple coherent revivals at the coarse-grained level, even though the full dynamics remain closed and unitary \cite{GoldsteinLebowitzTumulkaZanghi2006,PopescuShortWinter2006}.

\begin{figure}[t]
    \centering
    \includegraphics[width=0.65\linewidth]{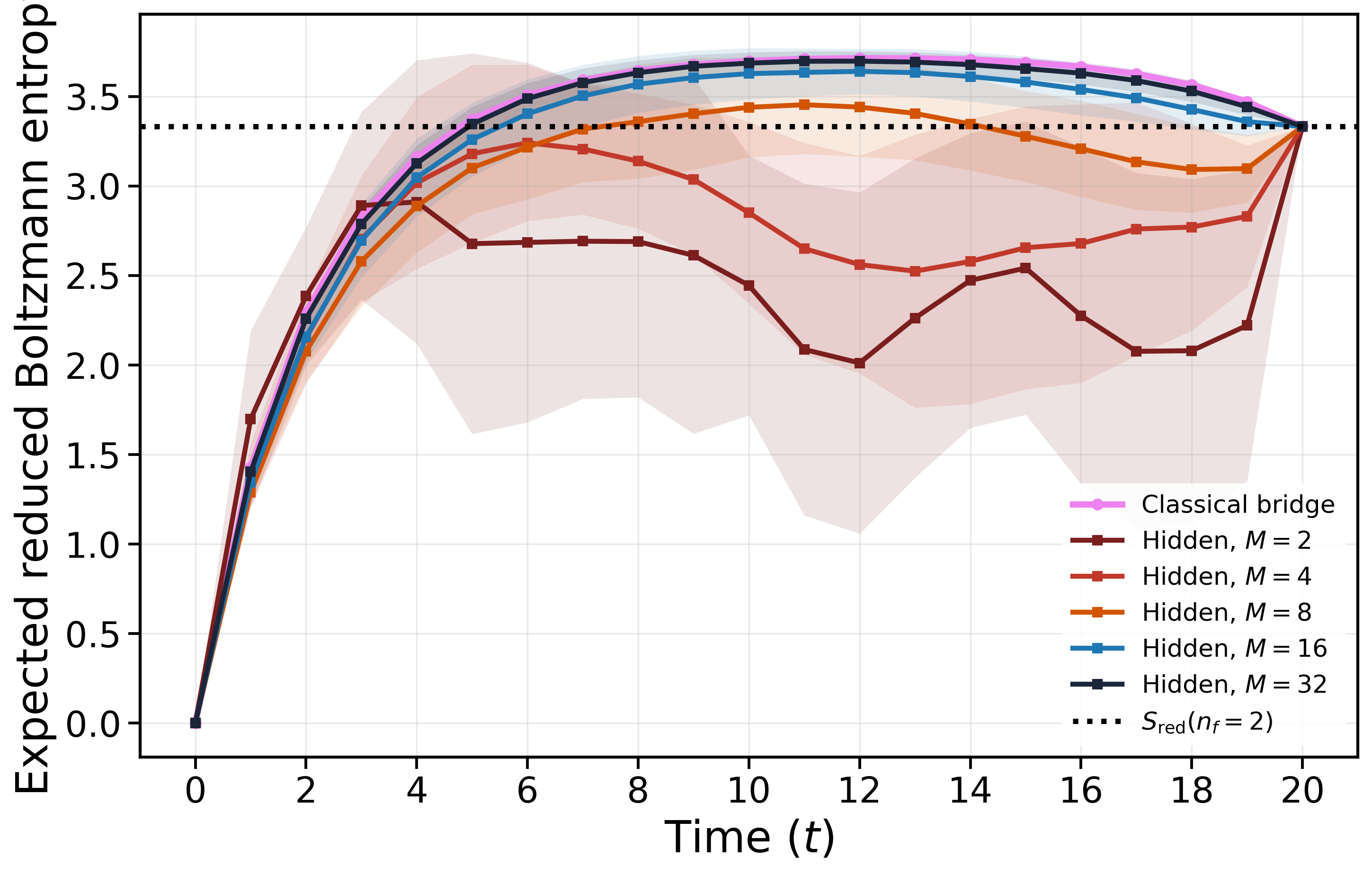}
\caption{ Hidden-microstate bridge entropy profiles for increasing internal Hilbert-space dimension \(M\), compared with the classical Boltzmann Bridge reference for \(N=8\), \(\alpha=0.1\), \(t_f=20\), and \(n_f=2\). The plotted quantity is the bridge-conditioned expected reduced Boltzmann entropy \(E_{Q,\mathrm{red}}^{(M)}[S_{\mathrm{red}}(t)\mid a,b]\). For small \(M\), especially \(M=2\) and \(M=4\), coherent finite-dimensional fluctuations lead to visible deviations from the classical bridge. As \(M\) increases, the profile becomes smoother and approaches the classical two-time-conditioned entropy curve. Solid curves show ensemble averages over 20 random internal unitaries, and shaded regions indicate the standard deviation across the ensemble. }
    \label{fig:M_switching}
\end{figure}

Figure~\ref{fig:M_switching} shows the hidden-microstate bridge entropy profile for increasing values of the internal dimension \(M\), with the classical Boltzmann Bridge shown as a reference. For small values of \(M\), especially \(M=2\) and \(M=4\), the curves remain strongly sample-dependent and deviate from the classical bridge. As \(M\) increases, the hidden internal degrees of freedom produce stronger coarse-grained phase scrambling, and the entropy profile moves progressively toward the classical result. The approach is already visible for \(M=8\), becomes strong for \(M=16\), and is nearly complete for \(M=32\). This supports the interpretation that classical-like two-time-conditioned bridge behavior can emerge statistically from closed quantum dynamics when sufficiently rich unresolved internal structure is present.

\subsection{Parameter-space classification with Random Forests}
\label{subsec:random_forest_parameter_search}

In the previous subsection, we showed, at fixed remaining parameters, that the internal dimension $M$ controls the approach to classical-like bridge behavior. We now ask whether a model-free algorithm, given no knowledge of the bridge construction, can independently identify the same parameters that the analytic formulation predicts should govern the transition, and how it weights them. To this end, and to further characterize the boundary between classical-like and revival-dominated behavior across a broader parameter space, we performed an exploratory machine-learning analysis. The purpose was not to establish a sharp phase boundary, but to test whether the bridge regime can be predicted from model parameters and simple spectral features of the hidden internal unitary dynamics. Here we used a Random Forest classifier, an ensemble method built from decision trees trained on random subsets of the data and features. Each tree partitions the feature space according to splits that reduce class impurity, and the final classification is obtained by majority vote across the individual trees~\cite{breiman2001}.

The dataset was generated using the parameter values summarized in Table~\ref{tab:ml_parameters_features}. We fixed \(N=20\) and \(t_f=20\), and varied \(\alpha\), \(n_f\), and \(M\) over the values listed in the table. For each fixed parameter tuple \((N,\alpha,t_f,n_f,M)\), we generated \(200\) independent realizations of the hidden internal dynamics by sampling independent Haar-random matrices \(U_R\) and \(U_L\) \cite{Mezzadri2007}. Each realization produced one hidden-microstate bridge curve, which was compared with the corresponding classical Boltzmann Bridge. This procedure produced
\begin{equation}
7\times 4\times 7\times 200
=
39200
\label{eq:ml_sample_count}
\end{equation}
data samples.

The target label was defined using the sign-agreement score \(C_\sigma\), which measures the agreement between the hidden-microstate bridge sign pattern and the corresponding classical bridge sign pattern. We define
\begin{equation}
C_\sigma
=
\frac{1}{|T|}
\sum_{t\in T}
\mathbf{1}
\left[
\sigma_Q^{(M)}(t)
=
\sigma_{\mathrm{cl}}(t)
\right],
\label{eq:sign_agreement_score}
\end{equation}
where \(\mathbf{1}[\cdot]\) is the indicator function. In this analysis we use
\begin{equation}
T
=
\{2,4,6,8,10,12,14,16,18\}.
\label{eq:ml_time_grid}
\end{equation}
Samples with \(C_\sigma\geq 0.8\) were labeled as classical-like, while samples with \(C_\sigma<0.8\) were labeled as revival-dominated or mixed. The classifier was trained on \(80\%\) of the dataset, and the remaining \(20\%\) was reserved for testing. This random split tests interpolation within the sampled parameter grid. 

\begin{table}[t]
\centering
\caption{Parameter grid, Random Forest input features, and target quantities used in the machine-learning analysis.}
\label{tab:ml_parameters_features}
\begin{tabular}{lll}
\hline
Quantity & Values or definition & Role \\
\hline
\(N\)
& \(20\)
& Fixed dataset parameter \\

\(t_f\)
& \(20\)
& Fixed dataset parameter \\

\(\alpha\)
& \(0.05,\;0.08,\;0.10,\;0.12,\;0.15,\;0.18,\;0.20\)
& Dataset parameter and RF feature \\

\(n_f\)
& \(1,\;3,\;5,\;8\)
& Dataset parameter \\

\(n_f/N\)
& \(0.05,\;0.15,\;0.25,\;0.40\)
& Derived RF feature \\

\(M\)
& \(2,\;4,\;8,\;12,\;16,\;24,\;32\)
& Dataset parameter and RF feature \\

\(\alpha t_f\)
& \(1.0,\;1.6,\;2.0,\;2.4,\;3.0,\;3.6,\;4.0\)
& Derived RF feature \\

\(\kappa_R^{(\tau)}\)
& \(\left|\operatorname{Tr}(U_R^\tau)\right|^2/M^2,\quad \tau=1,2,3\)
& Internal-unitary spectral RF feature \\

\(\kappa_L^{(\tau)}\)
& \(\left|\operatorname{Tr}(U_L^\tau)\right|^2/M^2,\quad \tau=1,2,3\)
& Internal-unitary spectral RF feature \\

\(\kappa_{RL}\)
& \(\left|\operatorname{Tr}(U_R^\dagger U_L)\right|^2/M^2\)
& Relative spectral-overlap RF feature \\

\(C_\sigma\)
& \(\displaystyle C_\sigma=\frac{1}{|T|}\sum_{t\in T}\mathbf{1}\left[\sigma_Q^{(M)}(t)=\sigma_{\mathrm{cl}}(t)\right]\)
& Target diagnostic, not an RF input feature \\

Label
& \(1\) if \(C_\sigma\geq 0.8\), otherwise \(0\)
& RF classification target \\

Samples per combination
& \(200\)
& Random realizations of \(U_R,U_L\) \\
\hline
\end{tabular}
\end{table}

\begin{figure}[H]
    \centering
    \includegraphics[width=0.35\linewidth]{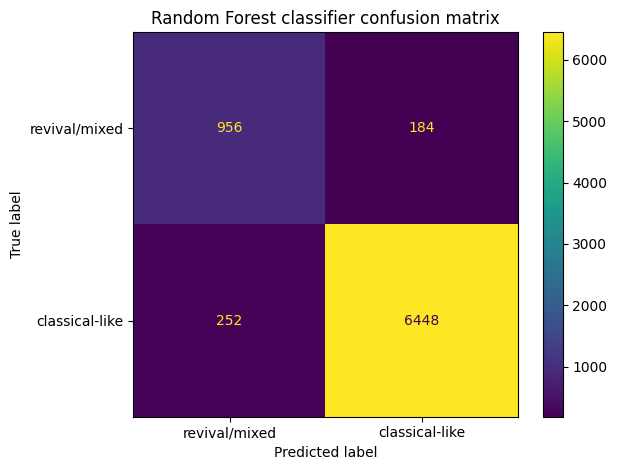}
    \includegraphics[width=0.50\linewidth]{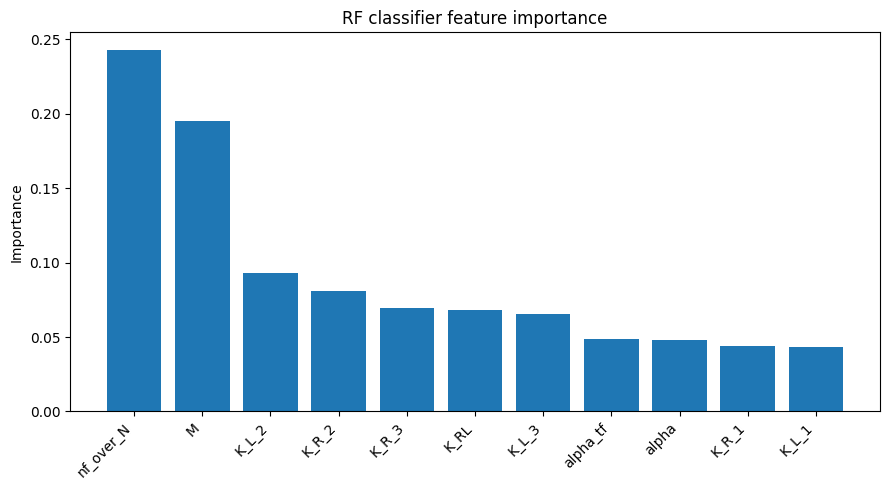}
    \caption{
    Random Forest classification of hidden-microstate bridge behavior. Left panel: confusion matrix on the held-out test set, comparing the true labels with the predicted labels for revival-dominated/mixed and classical-like samples. Right panel: feature-importance scores. The endpoint ratio \(n_f/N\) and the hidden-state dimension \(M\) are the most informative features, while the spectral recurrence features also contribute to identifying the transition region.
    }
    \label{fig:confusion-matrix}
\end{figure}

Figure~\ref{fig:confusion-matrix} summarizes the performance of the Random Forest classifier on the held-out test set. The left panel shows the confusion matrix for the two regimes: revival-dominated/mixed and classical-like. The classifier achieves an overall accuracy of approximately \(94\%\) and a macro-averaged F1 score of about \(89\%\). We report that for many parameter choices, particularly at larger hidden-state multiplicity \(M\), the bridge is more frequently classified as classical-like. Thus, the number of classical-like samples is larger than the number of revival-dominated/mixed samples. Despite this imbalance, the classifier retains useful performance, as indicated by accuracy and the F1 score, suggesting that it has learned a nontrivial separation between the two regimes rather than simply predicting the dominant class.

The right panel of Fig.~\ref{fig:confusion-matrix} shows the corresponding feature-importance scores. The largest importance is associated with the endpoint ratio \(n_f/N\) and the hidden-state multiplicity \(M\). This indicates that the transition between revival-dominated/mixed and classical-like behavior is not controlled by \(M\) alone. Instead, the classification boundary depends on the combined effect of the endpoint constraint and the hidden-microstate structure. In this interpretation, \(M\) controls the size of the unresolved internal Hilbert space and the tendency toward classical-like coarse-grained behavior, while \(n_f/N\) controls the strength and location of the imposed final macrocondition. The spectral recurrence features \(\kappa_R^{(\tau)}\), \(\kappa_L^{(\tau)}\), and \(\kappa_{RL}\) also contribute, showing that residual coherent structure in the hidden dynamics remains relevant near the transition region.

For visualization, we define the combined spectral-recurrence feature
\begin{equation}
\kappa_{\mathrm{sum}}
=
\sum_{\tau=1}^{3}
\left(
\kappa_R^{(\tau)}
+
\kappa_L^{(\tau)}
\right)
+
\kappa_{RL}.
\label{eq:kappa_sum}
\end{equation}

\begin{figure}[H]
    \centering
    \includegraphics[width=0.55\linewidth]{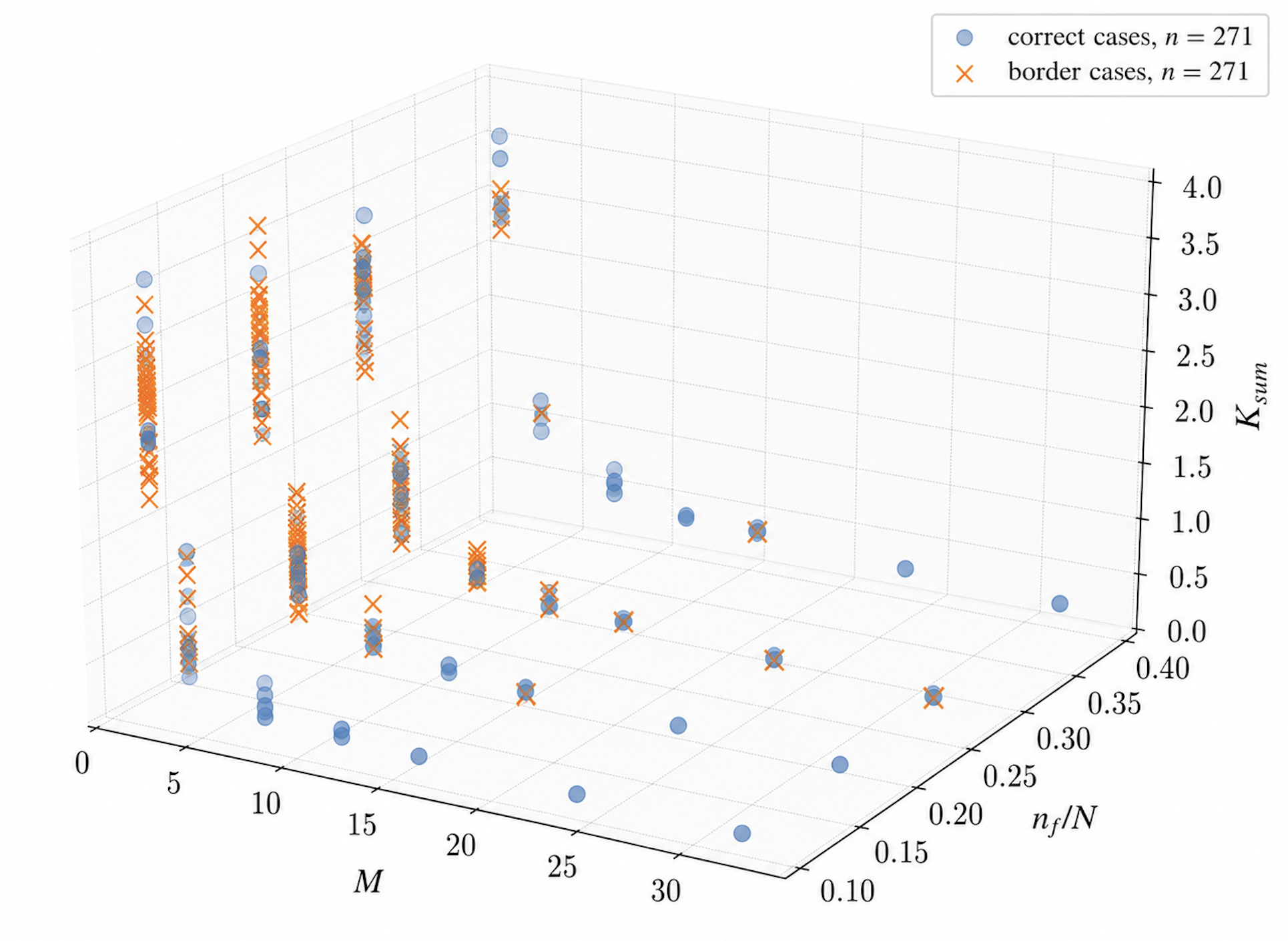}
    \caption{
    Border cases (orange) that are misclassified and correctly classified samples (blue) in the \((M,n_f/N,\kappa_{\mathrm{sum}})\) feature space. Border cases are samples with sign-agreement scores close to the empirical ambiguity threshold \(C_\sigma^\ast\simeq0.78\), near the nominal labeling threshold \(C_\sigma=0.8\). These samples are concentrated mainly at low-to-moderate \(M\) and larger spectral-recurrence values, where hidden internal dynamics retain stronger coherent structure. At larger \(M\), the ambiguity decreases and the bridge behavior becomes more robustly classical-like.
    }
    \label{fig:confusing}
\end{figure}

Figure~\ref{fig:confusing} shows the misclassified samples together with a
representative subset of correctly classified cases in the
\((M,n_f/N,\kappa_{\mathrm{sum}})\) feature space. The classification errors are
concentrated near the labeling threshold \(C_\sigma=0.8\), close to the
empirical ambiguity point \(C_\sigma^\ast\simeq 0.78\). Samples with diagnostic
scores well separated from this threshold are classified reliably, whereas
errors occur mainly in the region where the hard label assignment is
intrinsically ambiguous. This suggests that the classifier captures the
underlying regime structure and fails primarily at the definitional boundary
between classes, rather than exhibiting systematic errors throughout the
feature space.

The overlap is most pronounced at low-to-moderate values of \(M\), where \(\kappa_{\mathrm{sum}}\) is relatively large and the hidden dynamics retain stronger spectral recurrence. As \(M\) increases, \(\kappa_{\mathrm{sum}}\) typically decreases and the ambiguity is reduced, consistent with the system moving toward a more robust classical-like regime. The scatter plot also suggests that ambiguity depends on the endpoint ratio \(n_f/N\): larger endpoint fractions appear to produce fewer ambiguous cases, indicating that stronger endpoint conditioning can make the regime distinction more stable.

\section{Conclusion}
\label{sec:conclusion}

In this work, we formulated closed quantum analogues of the Boltzmann Bridge. The classical Boltzmann Bridge shows that an entropy history conditioned on both an initial low-entropy macrostate and a later macrostate can differ substantially from the usual past-only entropy curve. In particular, the bridge-conditioned entropy can rise above the entropy of the final macrostate at intermediate times and then decrease toward the imposed final endpoint. Our goal was to ask how this two-time-conditioned construction should be represented in a closed quantum system, where microscopic evolution is unitary and macrostates are represented by Hilbert-space subspaces rather than classical subsets of phase space.

The first step was to define a general quantum bridge construction using macro-subspace projectors, unitary time evolution, and Boltzmann entropy defined by the dimension of macroscopic sectors. This construction makes clear that the relevant entropy is not the global von Neumann entropy of the closed quantum state. Under unitary evolution, the global von Neumann entropy is conserved and therefore cannot describe the bridge entropy profile studied here. Instead, the appropriate quantity is the expected Boltzmann entropy over coarse-grained macroscopic sectors. The bridge probabilities should therefore be understood as probabilities for macro-projective, coarse-grained quantum histories.

We then studied the coherent chamber-qubit bridge, which is the most direct quantization of the classical two-box model. In this model, each particle carries only a two-state chamber degree of freedom, and the classical chamber bit is replaced by a qubit. Although this construction exactly preserves the same macroscopic chamber variable and the same Boltzmann entropy as the classical model, it does not reproduce classical bridge behavior. The reason is that the multi-step dynamics are coherent and unitary rather than Markovian. The resulting entropy profiles exhibit oscillations and revivals, and negative values of the bridge sign diagnostic occur only at isolated times rather than forming a stable classical-like bridge region. Thus, simply quantizing the chamber variable is not sufficient to recover the classical Boltzmann Bridge.

To go beyond this minimal model, we introduced the hidden-microstate bridge. In this construction, each chamber sector contains an internal Hilbert space of dimension \(M\). The chamber label remains the macroscopic observable, while the internal degrees of freedom remain unresolved. The full dynamics are still closed and unitary, but the hidden internal structure allows phase information to be redistributed within the coarse chamber sectors. This produces a much richer bridge propagator and gives a concrete mechanism by which classical-like coarse-grained behavior can emerge without adding external dissipation or stochasticity.

The numerical comparison supports this interpretation. For small internal dimension, such as \(M=2\) and \(M=4\), the hidden-microstate bridge remains strongly sample-dependent and does not reliably reproduce the classical sign structure. For intermediate dimension, such as \(M=8\), a classical-like negative bridge region begins to appear in the middle of the time interval. For larger internal dimension, especially \(M=16\) and \(M=32\), the hidden-microstate bridge increasingly agrees with the classical Boltzmann Bridge at the level of the reduced entropy profile and sign diagnostic. This shows that the classical pattern is not recovered by minimal quantization, but can arise statistically when the coarse macrostates contain sufficiently rich hidden internal structure.

The Random Forest analysis provided an additional parameter-space view of this transition. The classifier was able to distinguish classical-like bridge behavior from revival-dominated or mixed behavior with high accuracy on the held-out test set. The feature-importance results show that the transition is not controlled by the internal dimension \(M\) alone. The endpoint ratio \(n_f/N\), the effective switching scale, and simple spectral recurrence features of the internal unitaries also contribute. This supports the interpretation that classical-like bridge behavior depends on the combined effect of endpoint conditioning, hidden Hilbert-space dimension, and residual coherent structure in the internal dynamics.

Taken together, these results suggest a clear conclusion. Classical Boltzmann-bridge behavior does not appear simply by replacing a classical chamber bit with a coherent qubit. The minimal coherent quantum model remains dominated by unitary recurrence. Classical-like two-time-conditioned entropy behavior emerges only after the macroscopic chamber sectors are enlarged by unresolved internal degrees of freedom. In that sense, the hidden-microstate bridge provides a concrete route from closed unitary quantum dynamics to classical-looking two-time-conditioned entropy profiles.

Several directions remain open. The present work used a discrete two-chamber model with factorized one-particle dynamics, which allowed the bridge construction to be studied explicitly. Future work could extend the framework to interacting many-body systems, more general macroscopic observables, continuous-time dynamics, or spatially structured models. It would also be useful to compare the macro-projective bridge construction with decoherent-histories formulations in regimes where interference between coarse histories is dynamically suppressed. Finally, the machine-learning analysis could be expanded using grouped validation over unseen parameter tuples and richer spectral or dynamical features. These extensions would help clarify how general the hidden-microstate mechanism is and how broadly closed quantum systems can reproduce classical two-time-conditioned thermodynamic behavior under coarse-graining.

\section*{Acknowledgments}

The authors are grateful to Carlo Rovelli for introducing the Boltzmann Bridge problem and for many insightful discussions on the role of two-time conditioning in statistical mechanics. These conversations helped shape the motivation for the quantum extension developed in this work.

\appendix

\section{Explicit Dicke-amplitude formula for the coherent chamber-qubit bridge}
\label{app:dicke_amplitude_coherent_bridge}

This appendix gives the explicit Dicke-sector amplitude used in the coherent chamber-qubit bridge. In this model, each particle has two chamber states, \(|R\rangle\) and \(|L\rangle\), and the full Hilbert space is
\begin{equation}
\mathcal H
=
(\mathbb C^2)^{\otimes N}.
\label{eq:app_coherent_hilbert}
\end{equation}
The one-particle chamber rotation is
\begin{equation}
u(\theta)
=
\begin{pmatrix}
\cos\theta & -\sin\theta\\
\sin\theta & \cos\theta
\end{pmatrix},
\label{eq:app_one_particle_rotation}
\end{equation}
written in the ordered basis \((|R\rangle,|L\rangle)\). The full \(N\)-particle unitary is
\begin{equation}
U
=
u(\theta)^{\otimes N}.
\label{eq:app_many_particle_unitary}
\end{equation}
Since
\begin{equation}
u(\theta)^t
=
u(t\theta),
\label{eq:app_rotation_power}
\end{equation}
we define
\begin{equation}
c_t
=
\cos(t\theta),
\qquad
s_t
=
\sin(t\theta).
\label{eq:app_ct_st}
\end{equation}

Starting from the all-right state, the state at time \(t\) is
\begin{equation}
|\psi_t\rangle
=
(c_t|R\rangle+s_t|L\rangle)^{\otimes N}.
\label{eq:app_state_time_t}
\end{equation}
Let \(|D_n\rangle\) denote the normalized symmetric Dicke state with exactly \(n\) particles in the left chamber:
\begin{equation}
|D_n\rangle
=
\frac{1}{\sqrt{\binom{N}{n}}}
\sum_{\mathrm{wt}(x)=n}
|x\rangle .
\label{eq:app_dicke_state}
\end{equation}
Here the sum runs over all chamber strings \(x\in\{R,L\}^N\) containing exactly \(n\) copies of \(L\). Expanding Eq.~\eqref{eq:app_state_time_t} in this basis gives
\begin{equation}
|\psi_t\rangle
=
\sum_{n=0}^{N}
\sqrt{\binom{N}{n}}\,
c_t^{N-n}
s_t^n
|D_n\rangle .
\label{eq:app_dicke_expansion}
\end{equation}

The coherent bridge distribution depends on the Dicke-sector transition amplitude
\begin{equation}
A_{m,n}(\Delta)
=
\langle D_m|U^\Delta|D_n\rangle ,
\label{eq:app_dicke_amplitude_def}
\end{equation}
where
\begin{equation}
\Delta
=
t_f-t
\label{eq:app_delta_def}
\end{equation}
is the remaining number of time steps between the intermediate and final macro-conditioning. We also define
\begin{equation}
c_\Delta
=
\cos(\Delta\theta),
\qquad
s_\Delta
=
\sin(\Delta\theta).
\label{eq:app_cdelta_sdelta}
\end{equation}

For the rotation model, the amplitude from an intermediate Dicke sector \(n\) to a final Dicke sector \(m\) is
\begin{equation}
A_{m,n}(\Delta)
=
\sqrt{
\frac{\binom{N}{m}}{\binom{N}{n}}
}
\sum_{j=j_{\min}}^{j_{\max}}
(-1)^{n-j}
\binom{m}{j}
\binom{N-m}{n-j}
c_\Delta^{N-m-n+2j}
s_\Delta^{m+n-2j}.
\label{eq:app_dicke_amplitude_explicit}
\end{equation}
The summation limits are
\begin{equation}
j_{\min}
=
\max(0,n+m-N),
\qquad
j_{\max}
=
\min(n,m).
\label{eq:app_j_limits}
\end{equation}
The index \(j\) counts the number of particles that are in the left chamber both at the intermediate time and at the final time. The factor \(\binom{m}{j}\) chooses the particles that are left at both times, while \(\binom{N-m}{n-j}\) chooses the particles that are left at the intermediate time but right at the final time. The sign \((-1)^{n-j}\) comes from the convention in Eq.~\eqref{eq:app_one_particle_rotation}, where the amplitude for \(L\to R\) is \(-s_\Delta\).

For the bridge problem, the final sector is \(m=n_f\). Thus,
\begin{equation}
A_{n_f,n}(\Delta)
=
\sqrt{
\frac{\binom{N}{n_f}}{\binom{N}{n}}
}
\sum_{j=j_{\min}}^{j_{\max}}
(-1)^{n-j}
\binom{n_f}{j}
\binom{N-n_f}{n-j}
c_\Delta^{N-n_f-n+2j}
s_\Delta^{n_f+n-2j}.
\label{eq:app_dicke_amplitude_nf}
\end{equation}
In this equation, the limits are
\begin{equation}
j_{\min}
=
\max(0,n+n_f-N),
\qquad
j_{\max}
=
\min(n,n_f).
\label{eq:app_j_limits_nf}
\end{equation}

The unnormalized coherent bridge weight is
\begin{equation}
\widetilde p_Q^{\mathrm{coh}}(n,t;a,b)
=
\binom{N}{n}
c_t^{2(N-n)}
s_t^{2n}
\left|A_{n_f,n}(\Delta)\right|^2 .
\label{eq:app_bridge_weight_with_A}
\end{equation}
Therefore, the normalized bridge-conditioned distribution is
\begin{equation}
p_Q^{\mathrm{coh}}(n,t\mid a,b)
=
\frac{
\binom{N}{n}
c_t^{2(N-n)}
s_t^{2n}
\left|A_{n_f,n}(\Delta)\right|^2
}{
\sum_{r=0}^{N}
\binom{N}{r}
c_t^{2(N-r)}
s_t^{2r}
\left|A_{n_f,r}(\Delta)\right|^2
}.
\label{eq:app_bridge_distribution_with_A}
\end{equation}

For numerical evaluation, it is convenient to remove a common binomial factor that cancels between numerator and denominator. Define
\begin{equation}
F_{m,n}(\Delta)
=
\sum_{j=j_{\min}}^{j_{\max}}
(-1)^{n-j}
\binom{m}{j}
\binom{N-m}{n-j}
c_\Delta^{N-m-n+2j}
s_\Delta^{m+n-2j}.
\label{eq:app_F_def}
\end{equation}
Then
\begin{equation}
A_{m,n}(\Delta)
=
\sqrt{
\frac{\binom{N}{m}}{\binom{N}{n}}
}
F_{m,n}(\Delta).
\label{eq:app_A_in_terms_of_F}
\end{equation}
Substituting Eq.~\eqref{eq:app_A_in_terms_of_F} into Eq.~\eqref{eq:app_bridge_distribution_with_A}, the factor \(\binom{N}{n_f}\) cancels between numerator and denominator. Hence,
\begin{equation}
p_Q^{\mathrm{coh}}(n,t\mid a,b)
=
\frac{
c_t^{2(N-n)}
s_t^{2n}
\left|F_{n_f,n}(\Delta)\right|^2
}{
\sum_{r=0}^{N}
c_t^{2(N-r)}
s_t^{2r}
\left|F_{n_f,r}(\Delta)\right|^2
}.
\label{eq:app_bridge_distribution_F}
\end{equation}
For the real rotation matrix used in this model, \(F_{n_f,n}(\Delta)\) is real, and the modulus square can be replaced by an ordinary square:
\begin{equation}
p_Q^{\mathrm{coh}}(n,t\mid a,b)
=
\frac{
c_t^{2(N-n)}
s_t^{2n}
F_{n_f,n}(\Delta)^2
}{
\sum_{r=0}^{N}
c_t^{2(N-r)}
s_t^{2r}
F_{n_f,r}(\Delta)^2
}.
\label{eq:app_bridge_distribution_F_real}
\end{equation}

The coherent bridge entropy is then obtained from
\begin{equation}
E_Q^{\mathrm{coh}}[S_B(t)\mid a,b]
=
\sum_{n=0}^{N}
S_B(n)\,
p_Q^{\mathrm{coh}}(n,t\mid a,b),
\label{eq:app_bridge_entropy}
\end{equation}
with
\begin{equation}
S_B(n)
=
\log \binom{N}{n}.
\label{eq:app_coherent_entropy}
\end{equation}

Finally, if the coherent chamber-qubit model is calibrated to the classical one-step switching probability \(\alpha\), then
\begin{equation}
\sin^2\theta
=
\alpha,
\qquad
\theta
=
\arcsin\sqrt{\alpha}.
\label{eq:app_alpha_theta}
\end{equation}
Consequently,
\begin{equation}
c_t
=
\cos\left(t\arcsin\sqrt{\alpha}\right),
\qquad
s_t
=
\sin\left(t\arcsin\sqrt{\alpha}\right),
\label{eq:app_ct_st_alpha}
\end{equation}
and
\begin{equation}
c_\Delta
=
\cos\left((t_f-t)\arcsin\sqrt{\alpha}\right),
\qquad
s_\Delta
=
\sin\left((t_f-t)\arcsin\sqrt{\alpha}\right).
\label{eq:app_cdelta_sdelta_alpha}
\end{equation}
Thus, once \(N\), \(n_f\), \(t_f\), \(t\), and \(\alpha\) are specified, Eqs.~\eqref{eq:app_F_def}--\eqref{eq:app_bridge_entropy} determine the coherent chamber-qubit bridge.

\section{Explicit coarse propagator for the hidden-microstate bridge}
\label{app:hidden_microstate_K}

This appendix gives the explicit combinatorial form of the coarse propagator \(K_{m,n}^{(M)}(t,\Delta)\) used in the hidden-microstate bridge. In the main text, this propagator is defined by
\begin{equation}
K_{m,n}^{(M)}(t,\Delta)
=
\operatorname{Tr}
\left[
\Pi_m
U^\Delta
\omega_n(t)
(U^\dagger)^\Delta
\right],
\label{eq:app_hidden_K_trace}
\end{equation}
where the normalized post-measurement state is
\begin{equation}
\omega_n(t)
=
\frac{\Pi_n\rho_t\Pi_n}{p_Q^{(M)}(n,t\mid a)}.
\label{eq:app_hidden_omega}
\end{equation}
This expression is defined for sectors satisfying
\begin{equation}
p_Q^{(M)}(n,t\mid a)
>
0.
\label{eq:app_hidden_positive_sector}
\end{equation}

Let the one-particle \(t\)-step unitary be written as
\begin{equation}
u^t
=
\begin{pmatrix}
A_t & B_t\\
C_t & D_t
\end{pmatrix}.
\label{eq:app_hidden_t_step_unitary}
\end{equation}
The one-particle state at time \(t\), starting from the right-sector microcanonical state, is
\begin{equation}
\rho_t^{(1)}
=
u^t\frac{I_R}{M}(u^\dagger)^t .
\label{eq:app_hidden_one_particle_state}
\end{equation}
In chamber-block form this is
\begin{equation}
\rho_t^{(1)}
=
\begin{pmatrix}
R_t & X_t\\
Y_t & L_t
\end{pmatrix},
\label{eq:app_hidden_one_particle_blocks}
\end{equation}
where
\begin{equation}
R_t
=
\frac{1}{M}A_tA_t^\dagger,
\qquad
X_t
=
\frac{1}{M}A_tC_t^\dagger,
\qquad
Y_t
=
\frac{1}{M}C_tA_t^\dagger,
\qquad
L_t
=
\frac{1}{M}C_tC_t^\dagger .
\label{eq:app_hidden_RLXY}
\end{equation}
Here \(R_t\) and \(L_t\) are the diagonal chamber blocks, while \(X_t\) and \(Y_t\) are the off-diagonal chamber-coherence blocks.

Let \(P_R\) and \(P_L\) denote the one-particle projectors onto the right and left chamber sectors:
\begin{equation}
P_R
=
\sum_{\mu=1}^{M}
|R,\mu\rangle\langle R,\mu|,
\qquad
P_L
=
\sum_{\mu=1}^{M}
|L,\mu\rangle\langle L,\mu|.
\label{eq:app_hidden_PR_PL}
\end{equation}
The one-particle left probability at time \(t\) is
\begin{equation}
p_t
=
\operatorname{Tr}(L_t)
=
\frac{1}{M}
\operatorname{Tr}
\left(
C_tC_t^\dagger
\right).
\label{eq:app_hidden_pt}
\end{equation}
Similarly,
\begin{equation}
1-p_t
=
\operatorname{Tr}(R_t)
=
\frac{1}{M}
\operatorname{Tr}
\left(
A_tA_t^\dagger
\right).
\label{eq:app_hidden_one_minus_pt}
\end{equation}
Therefore, the probability of the intermediate macro-sector \(n\) is
\begin{equation}
p_Q^{(M)}(n,t\mid a)
=
\binom{N}{n}
p_t^n
(1-p_t)^{N-n}.
\label{eq:app_hidden_past_probability}
\end{equation}

For \(x,y,z\in\{R,L\}\), define the one-particle operators
\begin{equation}
O_{RR}(t)
=
R_t,
\qquad
O_{RL}(t)
=
X_t,
\qquad
O_{LR}(t)
=
Y_t,
\qquad
O_{LL}(t)
=
L_t .
\label{eq:app_hidden_Oxy_def}
\end{equation}
We then define the propagated scalar factors
\begin{equation}
q_{xy}^{z}(t,\Delta)
=
\operatorname{Tr}
\left[
P_z
u^\Delta
O_{xy}(t)
(u^\dagger)^\Delta
\right].
\label{eq:app_hidden_q_def}
\end{equation}
The labels \(x\) and \(y\) denote the chamber labels on the left and right sides of the intermediate density operator, while \(z\) denotes the final chamber label after evolution for the remaining time \(\Delta\). When \(x\neq y\), the operator \(O_{xy}(t)\) is off-diagonal in chamber space, and the corresponding factor \(q_{xy}^{z}(t,\Delta)\) can be complex. The full propagator \(K_{m,n}^{(M)}(t,\Delta)\), however, is real and nonnegative because it is defined by the trace expression in Eq.~\eqref{eq:app_hidden_K_trace}.

To write the many-particle propagator, introduce occupation numbers
\begin{equation}
N_{xy}^{z}
=
0,1,2,\ldots,
\qquad
x,y,z\in\{R,L\}.
\label{eq:app_hidden_occupations}
\end{equation}
Each \(N_{xy}^{z}\) counts how many particles contribute the factor \(q_{xy}^{z}(t,\Delta)\). Let \(\mathcal C_{m,n}\) denote the set of all nonnegative occupation numbers satisfying the following constraints:
\begin{equation}
\sum_{x,y,z\in\{R,L\}}
N_{xy}^{z}
=
N,
\label{eq:app_hidden_constraint_total}
\end{equation}
\begin{equation}
\sum_{y,z\in\{R,L\}}
N_{Ly}^{z}
=
n,
\label{eq:app_hidden_constraint_left_x}
\end{equation}
\begin{equation}
\sum_{x,z\in\{R,L\}}
N_{xL}^{z}
=
n,
\label{eq:app_hidden_constraint_left_y}
\end{equation}
and
\begin{equation}
\sum_{x,y\in\{R,L\}}
N_{xy}^{L}
=
m.
\label{eq:app_hidden_constraint_final}
\end{equation}
The first constraint fixes the total number of particles. The next two constraints enforce the intermediate projection onto the macro-sector with \(n\) left particles on the left and right sides of the density operator. The final constraint enforces the final macro-sector with \(m\) left particles.

The exact coarse propagator is then
\begin{equation}
K_{m,n}^{(M)}(t,\Delta)
=
\frac{1}{
\binom{N}{n}
p_t^n
(1-p_t)^{N-n}
}
\sum_{\{N_{xy}^{z}\}\in\mathcal C_{m,n}}
\frac{N!}{
\prod_{x,y,z\in\{R,L\}}
N_{xy}^{z}!
}
\prod_{x,y,z\in\{R,L\}}
\left[
q_{xy}^{z}(t,\Delta)
\right]^{N_{xy}^{z}} .
\label{eq:app_hidden_K_combinatorial}
\end{equation}
The denominator is exactly the probability \(p_Q^{(M)}(n,t\mid a)\) of the intermediate macro-sector. Thus, Eq.~\eqref{eq:app_hidden_K_combinatorial} is the normalized conditional propagator from the intermediate macro-sector \(n\) to the final macro-sector \(m\).

Although individual terms in the sum may be complex because of the off-diagonal factors \(q_{RL}^{z}\) and \(q_{LR}^{z}\), the final value of \(K_{m,n}^{(M)}(t,\Delta)\) is real and nonnegative. From the trace definition, it satisfies
\begin{equation}
K_{m,n}^{(M)}(t,\Delta)
\geq
0,
\qquad
\sum_{m=0}^{N}
K_{m,n}^{(M)}(t,\Delta)
=
1.
\label{eq:app_hidden_K_properties}
\end{equation}
The first property follows from positivity of the post-measurement state and the final macro-projector. The second follows from the completeness relation
\begin{equation}
\sum_{m=0}^{N}
\Pi_m
=
I.
\label{eq:app_hidden_projector_completeness}
\end{equation}

Finally, substituting \(K_{m,n}^{(M)}(t,\Delta)\) into the bridge formula gives
\begin{equation}
p_Q^{(M)}(n,t\mid a,b)
=
\frac{
p_Q^{(M)}(n,t\mid a)
K_{n_f,n}^{(M)}(t,\Delta)
}{
\sum_{\ell=0}^{N}
p_Q^{(M)}(\ell,t\mid a)
K_{n_f,\ell}^{(M)}(t,\Delta)
}.
\label{eq:app_hidden_bridge_distribution}
\end{equation}
This is the exact hidden-microstate bridge distribution used for numerical evaluation.

	\bibliographystyle{apsrev4-2}
	\bibliography{references}
\end{document}